\documentstyle[preprint,aps,eqsecnum]{revtex}
\begin{document}
\draft
%\preprint{dvi file made on \today}
\title{Kondo Effect in Flux Phases}
\author{Carlos R.~Cassanello and Eduardo Fradkin}
\bigskip
\address
{Loomis Laboratory of Physics and Materials Research Laboratory\\
University of Illinois at Urbana-Champaign\\
1110 W.Green St., Urbana, IL, 61801-3080}

\maketitle

\begin{abstract}
We consider a band of  fermions in two space dimensions with a flux
phase (relativistic) dispersion relation coupled to a local magnetic
impurity via an $ s-d$ interaction.
This model describes spinons of a flux phase and it is also
a qualitative model of the quasiparticles in a $d_{x^2-y^2}$
superconductor. We find a zero-temperature phase transition at a
finite coupling constant between a weak coupling unscreened impurity
state and a strong coupling regime with a
Kondo effect.  We use large-$N$ methods to study the phase transition
in this Kondo system away from  marginality. The Kondo energy scales
linearly with the distance to the transition . The zero-field magnetic
suceptibility at zero temperature diverges linearly. Similar behavior
is found in the $T$-matrix which shows a resonance at the Kondo scale.
However, in addition to this simple scaling,
we always find the presence of logarithmic corrections-to-scaling.
Such behavior is typical of systems at an upper critical dimension.
We derive an effective fermion model in one space dimension for this
problem. Unlike the usual Kondo problem, this system has an intrinsic
multichannel nature which follows from the spinor
structure of $2+1$-dimensional relativistic fermions.
\end{abstract}

\bigskip

\pacs{PACS numbers:~71.27.+a,75.20.Hr}

%\narrowtext

\section{Introduction}
\label{sec:intro}

It is by now well understood that the presence of a small
concentration of magnetic impurities into an otherwise non-magnetic
metallic host can affect dramatically the low-temperature properties
of the system. The prototype of
these interesting phenomena is the Kondo Effect.\cite{kondo}
At the single-impurity level, there is a non-perturbative crossover
between a Curie-Weiss law behavior (in which the impurity behaves like
a non-interacting localized magnetic moment) at high temperature, and
a strongly interacting regime in which the magnetic impurity and the
band electrons form a singlet ground state. In other words, in this
low-temperature (strong-coupling) regime, the band electrons conspire
to screen out the spin (and magnetic moment) of the impurity. At zero
temperature a similar crossover occurs as a function of an external
magnetic field. This picture has been developed by the concerted use
of the renormalization group~\cite{pwa,ayh}, ``exact" numerical
simulations and scaling~\cite{wilson}, exact solutions via the
Bethe-Ansatz~\cite{andrei,wiegmann} and large-$N$
expansions~\cite{read,piers}.

{}From the point of view of scaling, the Kondo problem is a typical
situation in which a
trivial fixed point, which describes band electrons decoupled from
the magnetic
impurity, is destabilized by a {\it marginally relevant} perturbation,
the coupling to
the magnetic impurity. This leads to an {\it asymptotically free}
renormalization
group flow with a $\beta$-function which is quadratic in the coupling
constant.
Marginal perturbations appear in critical systems at a {\it critical
dimension}.
The standard $s-d$ Kondo Hamiltonian is effectively a model of one
dimensional chiral
fermions coupled to a single magnetic impurity through the forward
scattering
channel~\cite{andrei}. Clearly, in this case we are at the
{\it lowest} critical
dimension. This is a direct consequence of the fact that the band
electrons have a
Fermi surface where the density of states is finite and essentially
constant.
Thus, the Kondo effect is ultimately due to the availability
of states in the electron band which can efficiently screen
the impurity spin no matter how weak the exchange coupling constant
may be.

Some time ago Withoff and Fradkin (WF)~\cite{withoff}
considered a genaralization of the Kondo problem
to systems in which the density of band electron states may actually
go to zero at
the Fermi energy.
They showed that if the density of states of the electron band
vanishes at the Fermi
energy as a positive power of the energy, the Kondo effect is
suppressed for small
values of the exchange constant and that the Kondo screening only
happens beyond a
critical value of this coupling. In fact, it is easy to see that the
exponent $r$
of the one particle density of states, $N(E) \propto | E-E_F|^{~r}$
plays a role here
quite analogous to the distance to the lower critical dimension
$d-d_c$ in
critical phenomena. WF showed, using a combination of a ``poor man's
scaling argument" and a large-$N$ limit, that at least for small
values of the
exponent $r$, this is the correct picture. Quite generally,
if the exponent $r>0$ this is a non-marginal Kondo system.

There are a number of systems of physical interest where this situation does
arise. A simple example are the fermionic excitations of a quantum
antiferromagnet in a {\it flux phase}~\cite{flux}.
More importantly, the normal state excitations of a $d$-wave superconductor
(with symmetry $d_{x^2-y^2}$) behave precisely in this
fashion~\cite{scalapino}.
In the vicinity of each {\it node} of the gap function (hence, the use of
the term  ``node" hereafter),
the dispersion relation for the normal quasiparticles is linear in the
momentum.
Thus, sufficiently close to the node,
the quasiparticles have an effective relativistic-like dispersion. In
the theory of superconductivity (isotropic or not)~\cite{schrieffer}
the dynamics of the quasiparticles is usually pictured in terms of Nambu
spinors. For the case of a d-wave superconductor, Nersesyan, Tsvelik and Wenger
~\cite{wenger} have shown that  this approach leads to
effective Hamiltonian for the quasiparticles which takes the form of a
massless Dirac Hamiltonian for each node of the gap, with the ``speed of light"
equal to the Fermi velocity.
(Naturally, relativistic massless Dirac fermions themselves always have this
property.)  In this paper we consider a
model which describes properly the coupling of flux-phase fermions to a
magnetic impurity. It turns out that this model can also be used to
describe the coupling of a local magnetic impurity to a d-wave
superconductor, including pair-breaking effects.
In a separate publication we will discuss in detail the problem
of a magnetic impurity in a $d$-wave superconductor in more
detail~\cite{borkowski,next}.

In this paper we reconsider the Kondo effect for non-marginal systems.
The model has fermions coupled locally to a magnetic impurity.
The fermions are assumed to obey a relativistic-like dispersion law and hence
a density of states vanishing linearly with the energy.
For simplicity we consider models with just one species of ``relativistic"
fermions. We will refer to them as  ``having a single node".
In particular we will discuss the case of an
impurity coupled to an electron band with a density of states that
vanishes
{\it linearly} with the distance to Fermi energy $E_F\equiv 0$.
This case was not examined by WF who found that the singularity
structure changed as soon as $r>1/2$.
We will show in this paper that at $r=1$ the additional singularities
conspire to give {\it simple scaling laws} modified
by {\it logarithmic corrections}~\cite{foot}. This picture is
strongly reminiscent
of a critical system at an {\it upper} critical dimension.

As in the conventional Kondo problem, here too we can construct an
effective  one-dimensional theory. However, when one carefully
reduces the $2+1$-dimensional fermions with a relativistic-like
energy-momentum dispersion to an effective $1+1$-dimensional model,
one finds that there are {\it at least} two angular momentum channels
that are {\it always} coupled. Thus, these are all {\sl inherently}
multichannel Kondo systems.
In particular, it is always the $\ell=0$ channel coupled either to
$\ell=1$ channel or to $\ell=-1$ channel.
Which pair of angular momentum channels are actually coupled depends
on the {\it Parity} of the node (or cone) to which the channels
belong. We find two equivalent ways
to represent the dimensionally reduced model. We have the freedom to
choose the effective $1+1$-dimensional fermions to have a local
kinetic energy and, hence, to behave like
conventional right movers. However, in this picture, the effective
interaction with the
impurity becomes non-local. The other, alternative picture, is to
have a local coupling
between right moving fermions which now have a non-local kinetic
energy.
In fact, due to phase space factors, the density of states of the
effective $1+1$-dimensional model goes to zero linearly with energy.
The linearity of the density of states reflects the fact that in the
original problem
the fermions move in two-space dimensions.
It is this feature what drives the system out of marginality and what
generates a
critical coupling constant below which the width of the
Kondo resonance vanishes.

In all cases of physical interest, {\it Parity} is an exact symmetry.
This means that the number of nodes (cones) is {\it even} and
that there should be as many cones with positive parity
as there are with negative parity. In the case of both the $d$-wave
superconductors and of the flux-phase fermions,
this property follows from the alternating signs of the gap function.
In contrast, if either parity or time reversal were broken,
all four cones would have the same properties under parity.
However, in that case there would always be a non-zero energy gap
on the entire Fermi surface. Thus, in general, the effective Kondo
hamiltonians always have an exact degeneracy (angular momentum
channels).
The spin symmetry is the usual $SU(2)$ spin rotation invariance
(which here we call {\it color}).
Here we will work with the $SU(N_c)$ generalization of this symmetry,
with the physical $N_c=2$.
The angular momentum degeneracy leads to and $SU(N_f)$ {\it flavor}
degeneracy.
For a problem with one node, $N_f=2$. When more than one node is
considered,
$N_f>2$. We will also consider impurity scattering amplitudes
which may change the angular momentum of the fermions.
The basic and simplest model is worked out in section \ref{sec:toy}.

In the second part of the paper we use large-$N$ methods to
investigate the behavior of these systems near the critical
coupling~\cite{read}. We find that there is still a Kondo scale once
we go over the critical coupling. However, the position and the width
of the resonance are not related anymore in the very simple way they
are in the usual Kondo effect. The details are worked out in
section \ref{sec:scaling}. The most salient feature of our results
is the presence of logarithmic corrections to simple scaling in all
quantities of physical interest, including
the zero-field paramagnetic suceptibility (at $T=0$), the Kondo
scale and the $T$-matrix for bulk fermions.

The large-$N_c$ theory that we present here suggests the following
scenario. For small values of the coupling constant $J$, the impurity is
free and effectively paramagnetic. In contrast, for large values of $J$,
the impurity is screened by the fermions. The critical coupling $J_c$ is
dimensionful and scales with the energy cutoff. However, the fact that these
systems are inherently multichannel Kondo systems suggests that for $J
\geq J_c$ the impurity is actually {\sl overscreened}~\cite{multichannel}.
In the language
of the Renormalization Group this requires  a flow with two finite fixed
points: one {\it infrared unstable} fixed point for the paramagnetic-Kondo
phase transition and an {\it infrared stable} to describe the
multichannel behavior of the Kondo phase. As in the
conventional Kondo problem, the large-$N_c$ theory can only describe the
formation of the Kondo singlet which in a non-marginal system can only
take place at finite coupling. In marginal Kondo systems multichannel behavior
is found as a next-to-leading order effect in the $1/N_c$
expansion or to third order in perturbative scaling~\cite{nozieres}.
In the systems of physical interest mentioned above where this model
applies, the number of channels $N_f$ is always larger than the number
of colors. In such a situation one can imagine that there exists a
critical number of channels such that both fixed points actually
coincide and below this critical number the transition dissapears
altogether~\cite{ingersent}. We should emphasize here that this physics
cannot be accessed by  a straightforward use of the large-$N_c$
expansion.

The paper is organized as follows. In section~\ref{sec:toy} we derive
the effective $1+1$-dimensional impurity models for systems of
fermions in two space dimensions with
relativistic-like dispersion (flux phase).
In section~\ref{sec:scaling} we study these models
in the large-$N$ limit and determine their critical behavior.
In section~\ref{sec:prop} we
calculate the propagator for the band fermions in the $N \to \infty$ limit and
use it to
derive the $T$-matrix. In section~\ref{sec:genimp} we consider a model with the
most
general type of scattering process for fermions with only one node coupled to a
single
impurity and determine the phase diagram. Section \ref{sec:conc} is devoted
to the conclusions and to
the discussions of the similarities and differences between
the problem we  discuss here and the
conventional Kondo problem.
In Appendix A we sketch the calculations of a few integrals.

\section{Toy model: s-d Hamiltonian for a linear density of states}
\label{sec:toy}

In this section we want to study the s-d Hamiltonian a
la Read and Newns \cite{read}, with the difference that we are not
assuming a constant density of states for the conduction fermions.
We are coupling an f-impurity to a bath of ``relativistic" electrons,
{\it ie.,\/} electrons with a linear dispersion relation. We assume the
chemical potential for the electrons (the Fermi energy) is zero. This
is a very special case but it will turn out to be quite important  for the
case of a d-wave superconductor as it will be discussed in the following
sections.

We start with a two-dimensional free fermion
Hamiltonian with a linear spectrum and a
Fermi velocity $v_F$ which may correspond to the linearization of a
fermion band of width $2D$, where $D$ is an energy cutoff. We will consider a
 model of band fermions with $N_c=N$ spin components with $N=2$
for the physically relevant
case of spin $SU(2)$. In general, this model will have a global $SU(N_c)$
``color" (or spin)
symmetry.
This model has a well defined large-$N$ limit. The free part of the Hamiltonian
is
\begin{equation}
H_0 =
\int \frac{d^2 p}{(2\pi)^2}
\Psi^{\dagger}_{\sigma i} (p)
\left(
v_F {\vec p}\cdot{\vec \alpha}
\right)_{ij}
\Psi_{\sigma j}(p)
\label{eq:h0}
\end{equation}
where $\alpha_1$ and $\alpha_2$ are Pauli matrices, so that
\begin{equation}
{\vec p}\cdot{\vec \alpha} = p_1 \alpha_1 + p_2 \alpha_2
=
p
\left(
\matrix{ 0 & e^{-i\theta}\cr
             e^{i\theta} & 0 \cr}
\right)
\label{eq:hdirac}
\end{equation}
Eq.(\ref{eq:hdirac}) can be diagonalized by expanding the fields into
a linear combination of spinor eigenstates. Let's call them $u_{\pm}$.
This eigenmodes satisfy
\begin{equation}
v_F {\vec p}\cdot{\vec \alpha}\; u_{\pm} = \pm v_F|p| u_{\pm}
\end{equation}
A particular (and convenient) choice of eigenstates is given by
\begin{equation}
u_{\pm}^{(1)} = \frac{1}{\sqrt{2}}\kern2cm {\rm and} \kern2cm
u_{\pm}^{(2)} = \pm \frac{e^{i\theta}}{\sqrt{2}}
\end{equation}
The fields $\Psi$'s carry a spin index $\sigma$ apart from the spinor index
$i$ and $j$ which we ignore at this point but we shall put
it back in when dealing with the interaction with the impurity term.
\begin{equation}
\Psi_{i} (p) = \sum_{\lambda = \pm}^{} u_{i}^{\lambda} (p) \zeta_{\lambda}(p)
\end{equation}
so that
\begin{eqnarray}
\Psi_1(p) & = & \frac{1}{\sqrt{2}} \left[ \zeta_{+}(p) + \zeta_{-}(p) \right]
\nonumber \\
\Psi_2(p) & = & \frac{1}{\sqrt{2}} \left[ \zeta_{+}(p) - \zeta_{-}(p) \right]
e^{i\theta}
\end{eqnarray}
Now we can expand the $\zeta$ fields in a basis of angular momentum
eigenmodes
\begin{equation}
\zeta_{\pm} (p)
=
\sum_{m=-\infty}^{\infty} e^{im\theta}  \zeta_{\pm , m} (p)
\end{equation}
where $m$ is an integer and $\pm$ indicates the positive/negative
energy spinor eigenmodes of the Dirac equation given by Eq.(\ref{eq:h0}).
We then have
\begin{equation}
H_0
=
\sum_{m=-\infty}^{\infty}
\int_{0}^{\infty}
\frac{p dp}{2\pi}  v_F |p|
\left[
\zeta^{\dagger}_{+ , m }(p) \zeta_{+ , m } (p)
-
\zeta^{\dagger}_{- , m }(p) \zeta_{- , m } (p)
\right]
\label{eq:ke}
\end{equation}
In real space we have
\begin{equation}
\Psi_{i} (\vec r) =
\int_{0}^{\infty} \frac{pdp}{2\pi} \sum_{\lambda=\pm}^{}
\int_{0}^{2\pi} \frac{d\theta}{2\pi}
\sum_{m=-\infty}^{\infty} u_{i}^{\lambda} (\theta) e^{i{\vec p}\cdot{\vec r}}
e^{im\theta} \zeta_{m , \lambda}(p)
\label{eq:psir}
\end{equation}
To reduce to an equivalent one-dimensional problem we use polar coordinates
where
$\vec p \equiv (p,\theta)$, $\vec r \equiv (r,\phi)$ and
$\vec p \cdot \vec r = pr \cos(\theta-\phi)$.
By making use of the expansion
\begin{equation}
e^{iz\cos\varphi} = \sum_{n=-\infty}^{\infty} i^n  J_n (z) e^{in\varphi}
\end{equation}
where $J_n (z)$ are the Bessel functions. They satisfy the relations
\begin{equation}
\int_0^{2\pi} \frac{d\theta}{2\pi}
e^{ipr\cos(\theta-\phi) \ + \ i m \theta} =
i^m { J}_m(pr) e^{i m\phi}
\kern1cm{\rm since}\kern1cm
{ J}_{-n}(z) = (-1)^n { J}_n (z)
\end{equation}
Using these in Eq.(\ref{eq:psir}) one gets
\begin{eqnarray}
\Psi_1(r, \phi)
&=&
\int_0^{\infty} \frac{pdp}{2\pi} \frac{1}{\sqrt{2}}
\sum_{m=-\infty}^{\infty}
i^m { J}_m(pr) \ e^{-im\phi}
\left[ \zeta_{m , +}(p) + \zeta_{m , -}(p) \right]
\nonumber \\
\Psi_2(r, \phi)
&=&
\int_0^{\infty} \frac{pdp}{2\pi} \frac{1}{\sqrt{2}}
\sum_{m=-\infty}^{\infty}
i^{m+1} { J}_{m+1}(pr) \ e^{-i(m+1)\phi}
\left[ \zeta_{m , +}(p) - \zeta_{m , -}(p) \right]
\end{eqnarray}
At the impurity site, {\it \/ i.e.,} for $r\to 0$, ${J}_0(0) =1$ and
${ J}_m(0) = 0$ for $m\neq 0$. So, we are left with
\begin{eqnarray}
\Psi_1(0)
&=&
\int_0^{\infty} \frac{pdp}{2\pi} \frac{1}{\sqrt{2}}
\left[ \zeta_{0 , +}(p) + \zeta_{0 , -}(p) \right]
\nonumber \\
\Psi_2(0)
&=&
\int_0^{\infty} \frac{pdp}{2\pi} \frac{1}{\sqrt{2}}
\left[ \zeta_{-1,  +}(p) - \zeta_{-1,  -}(p) \right]
\label{eq:Psi0}
\end{eqnarray}
This is telling us that the effect of having a cone-like dispersion relation
for the fermions instead of a flat band will not only reduce the available
density of states but it will also induce an angular momentum mode mixing.
Even though we may regard the impurity as having a $\delta$-function spatial
form factor as in the case of the usual Kondo effect, the coupling to the
fermions will no longer be restricted only to the s-wave, or the spherically
symmetric angular momentum channel. This feature will become evident in the
following step. We want to rewrite the theory by introducing effective
one-dimensional fermion operators. We use the radial component of
incoming and outgoing waves. We find two ``flavors" of right-movers given
by
\begin{equation}
\begin{array}{l l l l}
c_{1}(p) \equiv \ \sqrt{|p|} \ \zeta_{+ , 0}(|p|) ; & &
c_{2}(p) \equiv \ \sqrt{|p|} \ \zeta_{+ , -1}(|p|) ;  & {\rm for} \ \ p > 0
\\
 c_{1}(p) \equiv \ \sqrt{|p|} \ \zeta_{- , 0}(|p|) ; & &
c_{2}(p) \equiv - \ \sqrt{|p|} \ \zeta_{- , -1}(|p|) ; & {\rm for} \ \ p < 0
\label{eq:map}
\end{array}
\end{equation}
In other words, $c_1$ are the fermion states with angular momentum
$\ell=0$ while $c_2$ are
the fermion states with angular momentum $\ell=-1$.
We summarize these mappings as
\begin{equation}
\Psi_i(0) \ = {\frac{1}{\sqrt 2}} \int_{-\infty}^{\infty}
\frac{dp}{2\pi} \sqrt{|p|}
\ c_i(p)
\label{eq:psi}
\end{equation}
For a system with an $SU(N_c)$ symmetry,
the s-d interaction can be written in the form
\begin{equation}
H_{\rm imp} = \Psi^{\dagger}_{\alpha} (0)
\tau^{\rm a}_{\alpha\beta} \Psi_{\beta}(0) S^{ a} \ \equiv \
\Psi_{\alpha}^{\dagger}(0) \tau^{ a}_{\alpha\beta}
\Psi_{\beta}(0) \ f^{\dagger}_{\gamma}  \ \tau^{a}_{\gamma\delta} f_{\delta}
\label{eq:s-d}
\end{equation}
where $f^{\dagger}_{\gamma} \tau^{a}_{\gamma\delta} f_{\delta}$
is the impurity spin coupled to the electrons at the real-space position
$x = 0$. For a system with an $SU(N_c)$ symmetry, $a=1,\ldots, N_c^2-1$.
We will use the standard representation of spin operators (here, the
generators of $SU(N_c)$ defined in terms of the fermion operators $f_\gamma$ as
 $f^{\dagger}_{\gamma}  \tau^{ a}_{\gamma\delta} f_{\delta}$.
If the constraint $f^{\dagger}_{\gamma}f_\gamma=1$ is satisfied,
the impurity is in the lowest (fundamental) representation of $SU(N_c)$.
Other representations can be constructed by changing the
``filling fraction" of the impurity spin. Notice that,
for $SU(2)$, the only possible value of the filling fraction is equal to one.

Once again we stress the fact that the angular momentum channels
given by $m=0$ and $m=-1$ couple to the impurity. This is an important
difference with respect to the case of constant density of states, where
only the $m=0$ channel couples to the impurity.
We shall single out these two channels from the free part of the
Hamiltonian to obtain an effective theory
by replacing Eq. (\ref{eq:Psi0}) into Eq. (\ref{eq:s-d})
\begin{eqnarray}
H_{\it eff}
 = & &
\int_{0}^{\infty} \frac{p \ dp}{2\pi} v_F \ |p|
\left[
\zeta^{\dagger}_{0, + }(p)\zeta_{0, + }(p) - \zeta^{\dagger}_{0, - }(p)
\zeta_{0, - }(p)
\right]
\nonumber \\
&+& \
\int_{0}^{\infty} \frac{p \ dp}{2\pi} v_F \ |p|
\left[
\zeta^{\dagger}_{-1, + }(p)\zeta_{-1, + }(p) - \zeta^{\dagger}_{-1, - }(p)
\zeta_{-1, - }(p)
\right]
\nonumber \\
&+&
\frac{g}{2}
\int_{0}^{\infty} \frac{p \ dp}{2\pi}
\int_{0}^{\infty} \frac{q \ dq}{2\pi}
\Big[
\zeta^{\dagger}_{0, + }(p) \tau^{ a} \zeta_{0, + } (q)
+
\zeta^{\dagger}_{0, - }(p) \tau^{ a} \zeta_{0, - } (q)
\nonumber \\
&  & \qquad \qquad+ \ \
\zeta^{\dagger}_{0 ,+ }(p) \tau^{ a} \zeta_{0, - } (q)
+
\zeta^{\dagger}_{0, - }(p) \tau^{ a} \zeta_{0, + } (q)
\Big] \ S^{ a}
\nonumber \\
& + &
\frac{g}{2}
\int_{0}^{\infty} \frac{p \ dp}{2\pi}
\int_{0}^{\infty} \frac{q \ dq}{2\pi}
\Big[
\zeta^{\dagger}_{-1, + }(p) \tau^{ a} \zeta_{-1, + } (q)
+
\zeta^{\dagger}_{-1, - }(p) \tau^{a} \zeta_{-1, - } (q)
\nonumber \\
&  & \qquad \qquad - \ \
\zeta^{\dagger}_{-1, + }(p) \tau^{a} \zeta_{-1 ,- } (q)
-
\zeta^{\dagger}_{-1, - }(p) \tau^{a} \zeta_{-1, + } (q)
\Big] \ S^{ a}
\end{eqnarray}
By using the correspondence between the $\zeta$'s and $c$'s defined above
we may write the Hamiltonian in a more compact form.
Since we found that this model contains
effectively two ``flavors" of right movers, we define a flavor index
$l=1, \ldots, N_f$ with
$N_f=2$. The system has an effective $SU(2)$ ``flavor" symmetry which
originates in the
unavoidable mixing of angular momentum waves by the impurity. Similarly,
we will write down
the spin index $\sigma=1, \ldots, N_c$ explicitly. The effective Hamiltonian is
(repeated indices are summed)
\begin{equation}
H_{\it eff}
=
\int_{-\infty}^{\infty}
\frac{dp}{2\pi}  E(p)
c^{\dagger}_{l \sigma}(p) c_{l\sigma}(p)
+
\frac{g}{2}
\left[
\int^{\infty}_{-\infty} \frac{dp}{2\pi} \sqrt{|p|} \ c^{\dagger }_{l \sigma}(p)
\right]
\tau^{a}_{\sigma \omega}
\left[
\int^{\infty}_{-\infty} \frac{dq}{2\pi} \sqrt{|q|} \ c^{\dagger}_{l \omega}(q)
\right]
f^{\dagger}_{\alpha} \tau^{a}_{\alpha\beta} f_{\beta}
\label{eq:heff}
\end{equation}
where
$a=1, \ldots, N_c^2-1$. The single particle excitation energy is $E(p) = v_F \
p$.
Eq. (\ref{eq:heff}) shows that the coupling  between the impurity spin and
the band electrons is momentum dependent.

We note here that there is an alternative representation of this model.
It may be obtained
by, instead of rescaling the fields by a factor of $\sqrt{|p|}$ as we have done
here,
defining a new momentum variable  $k \equiv p^2$
(an later extending $k$ to $(-\infty, +\infty)$).
In this form, the effective Hamiltonian becomes
\begin{equation}
H_{\it eff} =
\int_{-\infty}^{\infty}
\frac{dk}{2\pi}  E(k)
c^{\dagger}_{l \sigma}(k) c_{l \sigma}(k)
+
\frac{g}{2}
\left[
\int^{\infty}_{-\infty} \frac{dk}{2\pi} \ c^{\dagger}_{l \sigma}(k)
\right]
\tau^{a}_{\sigma \omega}
\left[
\int^{\infty}_{-\infty} \frac{dk '}{2\pi}
\ c^{\dagger}_{l \omega}(k ')
\right]
f^{\dagger}_{\alpha} \ \tau^{a}_{\alpha\beta} f_{\beta}
\label{eq:halt}
\end{equation}
In this case $E(k) \equiv {\rm sgn}(k) \; \sqrt{k} $.
In this representation, the interaction
becomes local at the expense of a non-local kinetic energy. In fact these two
representations are equivalent and are the only ones compatible
with local (anti)-commutation relations for the $\Psi$-fields.
This is the representation used
by Withoff and Fradkin~\cite{withoff}.
In this paper we will not use this representation of
the model.

We may now proceed to study this theory with a functional integral
formalism\cite{read}. The Lagrangian, in imaginary time, is
\begin{eqnarray}
{\cal L}(\tau)
& = &
\int^{\infty}_{-\infty}  \frac{dp}{2\pi} \  c^{\dagger}_{l \sigma}(p)
\left( \frac{\partial}{\partial \tau} \ + \ E(p)\right) \ c_{l \sigma} (p)
\nonumber \\
& + &
\frac{J_0}{N_c}
\left[
\int^{\infty}_{-\infty} \frac{dp}{2\pi} \sqrt{|p|} \ c^{\dagger}_{l \sigma}(p)
\right]
\tau^{a}_{\sigma\omega}
\left[
\int^{\infty}_{-\infty} \frac{dq}{2\pi} \sqrt{|q|} \ c_{l \omega} (q)
\right]
f^{\dagger}_{\alpha} \ \tau^{ a}_{\alpha\beta} \ f_{\beta}
\nonumber  \\
& & \qquad +   f^{\dagger}_{\sigma}
\left(
\frac{\partial}{\partial\tau}  -  {\hat h}_{\sigma}
\right) f_{\sigma}
+
\epsilon_f(\tau)
\left(
f^{\dagger}_{\sigma} \ f_{\sigma} - Q_f  \right)
\label{eq:lag}
\end{eqnarray}
In Eq.(\ref{eq:lag}) we have defined $J_0\equiv g(N_c/2)$. We have also
included a magnetic field $\hat h$ which we will choose to be a diagonal matrix
${\hat
h}_\sigma$ and it will be defined below.
The Lagrange multiplier field $\epsilon_f(\tau)$ has been introduced to
enforce the constraint of charge (occupancy) $Q_f$ at the impurity site.
In principle, the partition function separates into a sum of subsectors each
of which is characterized by an impurity occupancy $Q_f$.

Following the standard large-$N_c$ decoupling approach of Read and
Newns~\cite{read},
we write the spin operators in terms of fermions and find an effective
four Fermi interaction. This interaction can be written in a simple form by
making use
of the well known identity, which holds for the generators of $SU(N_c)$
\begin{equation}
\sum_{\rm a=1}^{N_c^2-1} \tau^{ a}_{\sigma\omega} \ \tau^{ a}_{\alpha\beta}
=
\ N_c \ \delta_{\sigma\beta} \delta_{\omega\alpha} \ - \
\delta_{\sigma\omega} \delta_{\alpha\beta}
\label{eq:suntrace}
\end{equation}
A Hubbard-Stratonovich (H-S) transformation is now introduced to decouple the
fermionic quartic term which arises from this expansion.
Up to an integration over the H-S fields $\varphi_l(\tau)$
and $\varphi_l^{*}(\tau)$, Eq.(\ref{eq:lag}) is equivalent to
\begin{eqnarray}
{\cal L}'(\tau)
& = &
\int_{-\infty}^{\infty} \frac{dp}{2\pi}
c^{\dagger}_{l \sigma} (p)
\left(
\frac{\partial}{\partial\tau} + E(p)
\right)
c_{l \sigma}(p)
 +
\int_{-\infty}^{\infty} \frac{dp}{2\pi} \sqrt{|p|}
\left(
\varphi^{*}_{l}(\tau)  f^{\dagger}_{\sigma}
c_{l \sigma}(p)   +
\varphi_{l}(\tau)
c^{\dagger}_{l \sigma} (p) f_{\sigma}
\right)
\nonumber \\
& + &
\frac{N_c}{J_0} \sum_{l=1}^{N_f}
|\varphi_{l}(\tau)|^2 \ + \
f^{\dagger}_{\sigma}
\left(
\frac{\partial}{\partial\tau} -
{\hat h}_\sigma + \epsilon_f(\tau)
\right) f_{\sigma}
\ - \ Q_f \epsilon_f(\tau)
\label{eq:h-slag}
\end{eqnarray}
Eq.(\ref{eq:h-slag}) can be rearranged, by field shifting and completing
squares, in the form
\begin{eqnarray}
{\cal L}'(\tau)
& = &
\int_{-\infty}^{\infty} \frac{dp}{2\pi}
\left[
c^{\dagger}_{l \sigma} (p)
+
\varphi^{*}_{l}(\tau) \ f^{\dagger}_{\sigma} (\tau)
\sqrt{|p|}
\left(
\frac{\partial}{\partial\tau}  +  E(p)
\right)^{-1}
\right] \
\left(
\frac{\partial}{\partial\tau}  +  E(p)
\right)
\nonumber \\
& &  \kern2.5in
\left[
c_{l \sigma} (p)
+
\left(
\frac{\partial}{\partial\tau}  +  E(p)
\right)^{-1}\kern-.1in
\sqrt{|p|}
\varphi_{l}(\tau) \ f_{\sigma} (\tau)
\right]
\nonumber \\
& + &
f^{\dagger}_{\sigma} (\tau)
\left[
\frac{\partial}{\partial\tau}
-  {\hat h}_{\sigma}  +
\epsilon_f(\tau)
-
\varphi^{*}_{l}(\tau)
\int_{-\infty}^{\infty} \frac{dp}{2\pi}
\ |p|
\left(
\frac{\partial}{\partial\tau}  +  E(p)
\right)^{-1}  \kern-.1in
\varphi_{l}(\tau)
\right]
f_{\sigma}
\nonumber \\
& & \kern2in + \
\frac{N_c}{J_0} \sum_{l=1}^{N_f}
|\varphi_{l}(\tau)|^2
 -   Q_f \epsilon_f(\tau)
\label{eq:q-lag}
\end{eqnarray}
We now obtain an effective theory for the fields $\epsilon_f$ and
$\varphi_l$ by integrating out the fermions.
The partition function is given by
\begin{eqnarray}
Z & = &
\int D f^{\dagger} \ D  f \ D c^{\dagger} \ D  c \ D \epsilon_f \ D \varphi \
D \varphi^{*}
\ \exp( -\int d\tau {\cal L}' (\tau))
\nonumber \\
& = &
Z_0 \int  D\varphi  D\varphi^{*}  D \epsilon_f
\ \exp(-S_{eff})
\end{eqnarray}
where
\[
Z_0 = \exp\left[ \ N_c N_f \ {\rm Tr}
\int \frac{dp}{2\pi}
\log
\left(
\frac{\partial}{\partial \tau} + E(p)
\right)
\right]
\]
is the partition function of free fermions.
The impurity part of the effective action is
\begin{eqnarray}
S_{eff}
& = &
- \sum_{{\sigma} = 1}^{N_c}
{\rm Tr}
\log
\left[
\frac{\partial}{\partial \tau}  -
{\hat h}_{\sigma} +
\epsilon_f  -
\sum_{l=1}^{N_f} \varphi^{*}_{l} (\tau)
\int \frac{dp}{2\pi} |p| \frac{1}{\partial_{\tau} +  E(p)}
\ \varphi_{l} (\tau)
\right]
\nonumber \\
&  & \kern2in \ + \
\int d\tau \left(
\frac{N_c}{J_0}  \left( \sum_{l=1}^{N_f} |\varphi_{l}|^2\right)
 -  Q_f  \epsilon_f
\right)
\label{eq:seff}
\end{eqnarray}
Here we stress that the model of physical interest has $N_c=2$ and $N_f=2$.

The effective action of Eq.~(\ref{eq:seff}) has the standard form of
reference~\cite{read}. The key difference here is the form of the free fermion
Green's function which in this problem has a relativistic form.
For the usual Kondo problem the magnetic impurity is coupled to system of band
electrons with a constant density of states at the Fermi surface.
In the model that we discuss here, the density of states of the effective
fermions (the
``right movers") is still constant but the interaction with the magnetic
impurity has
an explicit momentum dependence. This momentum dependence is such that the
effective
coupling at low momenta becomes arbitrarily small.
We will show below that up to a critical value of the coupling constant $J_0$
there
is no Kondo effect. This is in fact the result of reference~\cite{withoff}.
Notice that the momentum dependence of the interaction, is a direct consequence
of
the relativistic dispersion. Should any finite density of states arise, either
by
effects of a chemical potential or induced by disorder, a crossover to a
conventional
Kondo effect will occur. There is an important physical case in which a finite
density of states is precluded by reasons of symmetry, and the Fermi energy has
to be
locked at zero. This is the case of the d-wave superconductors which we will
discuss elsewhere~\cite{next}.

We now consider the $N_c \to \infty$ limit. Here, {\it large} $N_c$ means the
limit
in which the rank of the group of spin rotations becomes large instead of being
$SU(2)$.
To proceed with the $1/N_c$ expansion, we look first for static
solution for $\varphi_{l}(\tau)$. From the $SU(N_f)$ flavor symmetry of the
effective
Hamiltonian, there is a manifold of solutions which span the group $SU(N_f)$.
Clearly, all the solutions break $SU(N_f)$ {\it spontaneously}.
As in all impurity problems, it is impossible to break spontaneously a
continuous
symmetry of bulk fermions by coupling them to an impurity, which has a finite
Hilbert space.  Thus, this apparent spontaneous symmetry breaking is an
artifact of
the $N_c \to \infty$ limit. In fact, we expect that it will already be restored
by the leading $1/N_c$ correction. This is precisely what happens in the
large-$N$
approach to the conventional Kondo problem~\cite{read}. Thus,
quantities which exhibit this apparent spontaneous symmetry breaking will get
strongly
corrected already in the next order in $1/N_c$.

In what follows we will seek a static, symmetric, solution of the form
$\varphi_l(\tau)=\varphi_0$. We will now derive the form of the Saddle Point
equations
which will determine $\varphi$ as a  function of $J_0$ and of the
filling fraction of the impurity $Q_f$.

First we need to compute
\begin{equation}
g_{0}(i\omega)
=  -  \int \frac{dp}{2\pi} |p| \ \frac{1}{-i\omega  +  p}
\label{eq:g0}
\end{equation}
By working in imaginary frequency we  automatically get the time-ordered
expression form for Eq.(\ref{eq:seff}).
To compute Eq.(\ref{eq:g0}) we introduce a lorentzian
cutoff function $f_{\Lambda}(p)$
and extend the integration over $p$ to $\pm\infty$. We have
\begin{equation}
g_{0}(i\omega)
=
\ - \ \int_{-\infty}^{\infty} \frac{dp}{2\pi} \frac{|p|}{p \ - \ i\omega}
\left(
\frac{\Lambda^2}{p^2 \ + \ \Lambda^2}
\right)
\ = \
\frac{i\omega}{2\pi v_F^2}
\frac{\Lambda^2}{\Lambda^2 - \omega^2}
\log\left(\frac{\omega^2}{\Lambda^2}\right)
\label{eq:g00}
\end{equation}
We can go back to Eq.(\ref{eq:seff}) which now reads
\begin{equation}
S_{eff}
  = - N_c  \beta
\int_0^{\infty}
\frac{d\omega}{2\pi}
\log\left[
\epsilon_f^2 +  \omega^2\left( 1  -  N_f|\varphi_0|^2 \ f(\omega^2, \Lambda^2)
\right)^2 \right]
 +
\beta
\left( \frac{ N_c N_f|\varphi_0|^2}{J_0}  -  Q_f  \epsilon_f
\right)
\end{equation}
where
\begin{equation}
f\left(\frac{\omega^2}{ \Lambda^2}\right) \ = \
\frac{1}{2\pi v_F^2} \frac{1}{1 \ - \ (\omega/\Lambda)^2}
\log\left(\frac{\omega^2}{\Lambda^2}\right)
\end{equation}
and the magnetic field $h$ has been set to zero.
The saddle point equations are
\begin{equation}
N_c   \int_{0}^{\infty} \frac{d\omega}{2\pi}
\ \frac{2\epsilon_f}{\epsilon_f^2 \ + \ \omega^2
\left( 1- N_f|\varphi_0|^2 \ f(\omega^2, \Lambda^2)\right)^2}
\ +  \ Q_f
\ = \ 0
\label{eq:phasesft}
\end{equation}
where we should understand the integral as computed using some
convenient adiabatic cutoff.
The non-trivial solution for $\varphi_0$ is given by
\begin{equation}
\frac{1}{J_0}
= \ - \
\int_{0}^{\infty}
\frac{d\omega}{2\pi} \
\frac{2 \ \omega^2 \ f(\omega^2, \Lambda^2)
\left(1 - N_f|\varphi_0|^2 \ f(\omega^2, \Lambda^2)\right)}
{\epsilon_f^2 \ + \ \omega^2\left( 1 - N_f|\varphi_0|^2 \ f (\omega^2,
\Lambda^2)
\right)^2}
\label{eq:J0}
\end{equation}
It will prove useful to define the dimensionless variables
$x\equiv \frac{\omega}{\Lambda}$ and
$\nu\equiv
\frac{|\epsilon_f|}{\Lambda}$.
The saddle point equations Eq.(\ref{eq:phasesft}) and Eq.(\ref{eq:J0})
now read
\begin{equation}
\frac{Q_f}{N_c} \ = \
- \ {\rm sgn}(\epsilon_f) \
\int_0^{\infty} \frac{dx}{\pi} \
\frac{\nu}{\nu^2 \ + \ x^2
\left[
1 - \Delta \frac{\log x}{1-x^2}
\right]^2}
\end{equation}
and
\begin{equation}
\frac{1}{J_0} \ = \
- \ \frac{\Lambda}{(\pi v_F)^2}
\int_0^{\infty} dx \ \frac{\log x}{1-x^2} \left( 1 - \Delta
\ \frac{\log x}{1-x^2}\right) \
\frac{x^2}{\nu^2 + x^2\left(1 - \Delta\frac{\log x}{1-x^2}\right)^2}
\end{equation}
where $\Delta \ \equiv \ N_f|\varphi_0|^2 / \pi v_F^2$.

There exists a critical value for $J_0$, which we define as the value of $J_0$
at the point where $\varphi_0$ departs from zero, for vanishing
$\epsilon_f$\cite{withoff}. It is given by
\begin{equation}
\frac{1}{J_c}
=
 -  \int_{0}^{\infty}
\frac{d\omega}{\pi}
\  f
\left(\frac{\omega^2}{ \Lambda^2}\right)
\ = \ - \
\frac{\Lambda}{(\pi v_F)^2}
\int_0^{\infty}
\ \frac{dx}{1-x^2}
\ \log  x
\ =  \
\frac{\Lambda}{(\pi v_F)^2} \
\frac{\pi^2}{4}
\end{equation}
In the next section we use the SPE to extract the critical behavior of the
system
near $J_c$.

\section{Scaling and energy scales in the static approximation}
\label{sec:scaling}

In the previous section we obtained the saddle-point equations (SPE) for this
theory; however, these SPE have a singular behavior around the point
$\nu = 0$, $\Delta = 0$.
In particular, an expansion in powers of $\Delta$ for small $\Delta$ is not
possible for $\nu\to 0$. In this section we want to investigate in further
detail the behavior of the SPE and the scaling behavior of $\Delta$ as we
approach the critical point.
We will find that, opposite to the situation in the usual Kondo problem,
there are now two independent energy scales for $\Delta$ and $\nu$.

We go back to the SPE expressed in their original form
\begin{equation}
- \ \frac{Q_f}{N_c} \ {\rm sgn}(\epsilon_f)
=
\frac{1}{\pi}
\int_0^{\infty} dx \ \frac{\nu}{\nu^2 + x^2\left( 1 - \Delta \
\frac{\log x}{1-x^2}
\right)^2}
\label{eq:scaling1}
\end{equation}
and
\begin{equation}
\frac{1}{g_0} = \ - \ \int_0^{\infty}
dx \ \frac{\log x}{1-x^2} \left( 1 - \Delta \ \frac{\log x}{1-x^2}
\right)
\ \frac{x^2}{\nu^2 + x^2\left( 1 - \Delta \
\frac{\log x}{1-x^2}
\right)^2}
\label{eq:scaling2}
\end{equation}
where \ $\Delta \equiv N_f|\varphi_0|^2/\pi v_F^2$ \ and \ $g_0 \equiv
\Lambda J_0/ (\pi v_F)^2$.

In Eq.(\ref{eq:scaling1}), the integrand is the equal time propagator for the
impurity, with the integration variable $x$ being the imaginary frequency
scaled by the band electrons cutoff $\Lambda$.
This equation sets up a scale for a crossover between two different
behaviors, in much the same way as in the usual Kondo effect
the imaginary part of the band electrons Green function sets the
scale of the Kondo temperature.

In the case of the usual Kondo effect, the expression equivalent to
Eq.(\ref{eq:scaling1}) is
\begin{equation}
- \ \frac{Q_f}{N_c} \ {\rm sgn}(\epsilon_f) = \frac{1}{\pi}
\int_0^{\infty} dx \ \frac{\nu}{\nu^2 + (x +  \Delta)^2}
\label{eq:scaling3}
\end{equation}
The case $N_c = 2$ is, in a sense, a limiting case as can be seen from
Eq.(\ref{eq:scaling3}). On one hand, the r.~h.~s.~ of this equation is a
positive
function. This fact forces ${\rm sgn}(\epsilon_f)$ to be negative in order
to have a solution. However, even in this case one gets
\begin{equation}
Q_f \ \frac{\pi}{2} = \arctan \left( \frac{\Lambda}{|\epsilon_f|} \right)
\ - \ \arctan \left( \frac{\Delta}{|\epsilon_f|} \right)
\label{eq:scaling4}
\end{equation}
where $\Lambda$ is some electron band cutoff.
This suggests that for $Q_f =0$, the impurity level
$|\epsilon_f|$ has to approach zero.
However, if $Q_f = 1$, there is no solution,
since $\pi/2 \ - \ \arctan(\Lambda/|\epsilon_f|)$ is
a strictly positive number.
Now, if $N_c$ is an integer bigger than 2, and letting
$\Lambda \to \infty$, we find
\begin{equation}
\pi \left(\frac{1}{2} - \frac{Q_f}{N_c}\right)
= \arctan \left( \frac{\Delta}{|\epsilon_f|} \right)
\label{eq:scaling5}
\end{equation}
Here we can distinguish two different regimes: $Q_f/N_c \ \to \ 1/2$ which
corresponds to the case $\Delta << |\epsilon_f|$, and
$Q_f/N_c \ \to \ 0$ which corresponds to $|\epsilon_f| << \Delta$.

However, for the system being discussed here, the situation is a little
different and,
actually, more complex since, as it turns out, now there are two different
scales
involved. In principle there
is a scale set on $x$ by $\Delta$ at the point where
\begin{equation}
1 \approx - \Delta \ \frac{\log x}{1-x^2}
\qquad {\rm or,} \qquad
\frac{1}{\Delta} \approx - \ \frac{\log x}{1-x^2}
\label{eq:scaling6}
\end{equation}
If the value of the frequency (and therefore of $x$) is
small enough so that $x^2 \ll 1$, we can approximate
\begin{equation}
x \approx \exp \left( - \ \frac{1}{\Delta} \right)
\label{eq:scaling7}
\end{equation}
For $\Delta$ small enough, the approximation is consistent.
Working on this idea, we split Eq.(\ref{eq:scaling1}) as
\begin{eqnarray}
\frac{Q_f}{N_c}
& = &
\frac{1}{\pi} \int_0^{e^{-1/\Delta}}
dx \ \frac{\nu}{\nu^2 + x^2\left(1-\Delta \ \frac{\log x}{1-x^2}\right)^2}
\ + \
\frac{1}{\pi} \int_{e^{-1/\Delta}}^{\infty}
dx \ \frac{\nu}{\nu^2 + x^2\left(1-\Delta \ \frac{\log x}{1-x^2}\right)^2}
\nonumber \\
& \equiv &
I_1 + I_2
\label{eq:scaling9}
\end{eqnarray}
where we have used the fact that the only solution consistent with $Q_f/N_c
<1/2$ has
$\epsilon_f<0$.

The SPE for $\Delta$ may be treated using a similar approach.
Eq.(\ref{eq:scaling2}) can be split into $I_1'$ and $I_2'$, where the first one
is the corresponding integral up to $e^{-1/\Delta}$, and the second one takes
over from that point to infinity. Again we are interested in the small $\Delta$
and small $\nu$ regime.
The detailed computation of the integrals $I_1$, $I_2$, and $I_1'$, $I_2'$ is
given in
appendix \ref{sec:A}.

We will be interested in the following limiting cases:
\newcounter{letras}
\begin{list}
{\alph{letras}~)}{\usecounter{letras}}
\item
$ \frac{Q_f}{N_c} << \frac{1}{\pi} $: \\
In the regime  in which
$\nu << e^{-1/\Delta} << 1$, with
$\nu, \Delta << 1$, the
contribution from $I_2$ is neglegible and
$\frac{Q_f}{N_c} \approx \frac{\nu}{\pi} \ e^{1/\Delta}$ \  or
\ $ \nu \approx \pi \ \frac{Q_f}{N_c}  \ e^{-1/\Delta} << e^{-1/\Delta}$. \
Thus, in this regime, \ $ \frac{Q_f}{N_c} << \frac{1}{\pi} $.
In this limit,
the leading term from the other SPE
is going to be (see Appendix \ref{sec:A}, Eq.(\ref{eq:applast}))
\begin{equation}
\left( \frac{\pi^2}{4} - \frac{1}{g_0}
\right) \ \approx \
e^{-1/\Delta}
\left[ \left( \pi \frac{Q_f}{N_c} \right)^2 \frac{1}{\Delta}
\right]
\label{eq:nu3}
\end{equation}
This gives
\begin{equation}
\Delta \ \approx \
\frac{1}{
\log
\left[
\frac{
\left( \pi \frac{Q_f}{N_c}\right)^2
}{
\frac{\pi^2}{4} - \frac{1}{g_0}
}
\right]
} \ - \
\frac{
\log \log
\left[
\frac{
\left( \pi \frac{Q_f}{N_c}\right)^2
}{
\frac{\pi^2}{4} - \frac{1}{g_0}
}
\right]
}
{
\log^2
\left[
\frac{
\left( \pi \frac{Q_f}{N_c}\right)^2
}{
\frac{\pi^2}{4} - \frac{1}{g_0}
}
\right]
}
\ + \dots
\label{eq:delta2}
\end{equation}
and
\begin{equation}
\nu \ \approx \
\pi \frac{Q_f}{N_c} e^{-1/\Delta}
\ \approx \
\frac{\left( \frac{\pi^2}{4}-\frac{1}{g_0}\right)}
{\pi \frac{Q_f}{N_c}}
\ \frac{1}
{\log
\left[
\frac{
\left( \pi \frac{Q_f}{N_c}\right)^2
}{
\frac{\pi^2}{4} - \frac{1}{g_0}
}
\right]
}
\label{eq:nu2}
\end{equation}
\item
$\frac{ Q_f}{N_c} \approx \frac{1}{2}$:\\
We now consider the opposite regime,
$ e^{-1/\Delta}  <<  \nu  <<  1 $, where we obtain
\begin{equation}
\frac{ Q_f}{N_c} \approx \frac{1}{2}  -  \frac{3}{\pi \nu^3} \ e^{-3/\Delta}
\ \ {\rm and} \ \
\nu \ \approx \ \left(\frac{3}{\pi}\right)^{1/3}
\ \frac{e^{-1/\Delta}}{\left( - \frac{Q_f}{N_c} + \frac{1}{2}\right)^{1/3}}
\ >> \ e^{-1/\Delta} \ \ {\rm if} \ \ \frac{Q_f}{N_c} \ \to \ \frac{1}{2}.
\label{eq:scaling16}
\end{equation}
In this regime, clearly $\frac{Q_f}{N_c} \ \to \ \frac{1}{2}$ and the other
SPE gives
\begin{equation}
\frac{1}{g_0} \ \approx \
\frac{\pi^2}{4} \ + \ \frac{\pi}{2} \ \nu \log \nu
\label{eq:scaling22}
\end{equation}
We can get a solution for $\nu$ by iteration on Eq.(\ref{eq:scaling22}),
\begin{equation}
\nu \ \approx \
\frac{1}
{
\log \left[
\frac{\pi}{2\left(\frac{\pi^2}{4} - \frac{1}{g_0}\right)}
\right]
} \
\frac{2}{\pi}
\left(
\frac{\pi^2}{4}-\frac{1}{g_0}\right)
\label{eq:nu1}
\end{equation}
Using Eq.(\ref{eq:scaling16}) we get $\nu$ which, after being replaced into
Eq.(\ref{eq:scaling22}) gives a scaling form for $\Delta$ in this regime
\begin{equation}
 \frac{2}{\pi}\left( \frac{\pi^2}{4} -
\frac{1}{g_0}\right)
\left[\frac{\pi}{3}
\left(\frac{1}{2} - \frac{Q_f}{N_c}\right)
\right]^{1/3}
\approx
 \ e^{-1/\Delta}
\log \left[
\frac{\pi}{2\left(\frac{\pi^2}{4} - \frac{1}{g_0}\right)}
\right]
\label{eq:scaling23}
\end{equation}
Thus,
\end{list}
\begin{equation}
\Delta
 \approx
\left[
\log
\left[
\frac{1}{ \frac{2}{\pi} \left( \frac{\pi^2}{4} - \frac{1}{g_0} \right)
\left( \frac{\pi}{3} \left(\frac{1}{2}-\frac{Q_f}{N_c}\right)\right)^{1/3}}
\right]
\right]^{-1}\kern-5pt
\left(
1  -  \frac{\log \log \left[
\frac{2}{\pi} \left( \frac{\pi^2}{4} - \frac{1}{g_0} \right)
\right]^{-1}}{\log \left[
\frac{2}{\pi} \left( \frac{\pi^2}{4} - \frac{1}{g_0} \right)
\left(\frac{\pi}{3}\left(\frac{1}{2}-\frac{Q_f}{N_c}\right)\right)^{1/3}
\right]^{-1}}
\right)
\label{eq:delta1}
\end{equation}
Notice that, for the physical case $N_c=2$,
the two regimes $Q_f/N_c << 1$ and $Q_f\approx N_c /2$ are identical.
Only as $N_c$ grows large it becomes possible to distinguish
one case from the other. However, in the case $Q_f\approx N_c/2$
the magnitude of $\Delta$
appears to go to zero as $Q_f \to N_c/2^-$ even though the critical
coupling is independent of the value of $Q_f$. In contrast, in the case
$Q_f/N_c << 1$ such unwanted feature is not present. It is worth to
note here that, in spite of this apparent difficulty, we will show at the
end of this section that the zero-temperature suceptibility at zero field
has the same behavior in both regimes. Thus, this difficulty is not
physically relevant. For simplicity, we will use the case $Q_f/N_c << 1$
to extrapolate to the physically meaningful case of $N_c=2$.

We can use these results to derive the $\beta$-function for the coupling
constant $g_0$ in the limit $N_c \to \infty$.
In order to do this we may use Eq.(\ref{eq:scaling22}) for the regime in which
$\frac{Q_f}{N_c} \ \approx \ \frac{1}{2}$, or Eq.(\ref{eq:nu2}) for the case
$\frac{Q_f}{N_c} \ \to \ 0$.
Starting from Eq.(\ref{eq:nu3}) we may replace $\Delta$ in terms of $\nu$ to
obtain
\begin{equation}
\frac{\pi^2}{4} \ - \ \frac{1}{g_0}
\ \approx \ \alpha \ \nu\left( \log\alpha \ - \ \log\nu\right)
\label{eq:beta1}
\end{equation}
where \ $ \alpha \ \equiv \ \pi \frac{Q_f}{N_c}$.
The $\beta$-function keeps track of the flow of $g_0$ as the cutoff
$\Lambda$ is decreased from very large values (infinite bandwidth). Hence,
\begin{equation}
\beta \left( g_0 \right) \ \equiv  \
- \ \Lambda \ \frac{\partial \ g_0}{\partial \Lambda}
\label{eq:beta}
\end{equation}
In both limits $Q_f/N_c \to 0$ and $Q_f/N_c  \to 1/2$  we find
\begin{equation}
\beta \left( g_0 \right)
= \ - \ g_0 \ + \ {\frac{1}{g_c}} \ g_0^2
\label{eq:beta2}
\end{equation}
where $ {\frac{1}{g_c}} =  \frac{\pi^2}{4}$ .

We immediately see that there are two fixed points: (a) a stable
fixed point at \ $g_0 \ = \ 0$ and (b) an unstable fixed point at
$g_0=g_c$. The stable fixed point represents the weak coupling phase in
which the fermions are decoupled from the impurity and the impurity spin
is unscreened and there is no Kondo effect.
The non-trivial fixed point  at $g_0=g_c$ separates the weak coupling
phase from a strong coupling phase that, in principle, should exhibit a
Kondo effect.  This unstable fixed point can be regarded as the usual
marginally unstable fixed point of the standard Kondo problem, now pushed to a
finite value of the coupling constant by effect of the reduction
of the density of band states. Hence, in the present case, the interaction
has to become strong enough so as to overcome the effect of the depletion
in the number of states available in order to produce the Kondo screening.
Thus, for systems with a small
coupling to the impurity, there will be no Kondo screening and
we should expect a decoupled impurity behavior.

If $g_0 >g_c$, the system is in the Kondo screeening  phase.
We may define a {\it Kondo scale} $T_K$ for the regime $g_0>g_c$. From
Eq.(\ref{eq:nu3}) through Eq.(\ref{eq:nu2}) we see that,
in terms of the physical parameters of the theory we have
\begin{equation}
T_K \equiv |\epsilon_f| \approx
\pi \frac{N_c}{Q_f} \ v_F^2 \left( \frac{J_0 - J_c}{J_0 \ J_c}\right)~\left(
\log\left[
\left( 2 \frac{Q_f}{N_c}\right)^2  \left(
\frac{J_0}{J_0-J_c}
\right) \right]\right)^{-1}
\end{equation}
The results found here generalize the work of WF~\cite{withoff}.
In that work the case of a density of states which
vanishes linearly was not discussed on account that the structure of the
singularities was not smoothly connected with the case of a constant density of
states.
The results of this section show that the main
difference is the presence of logarithmic corrections to scaling in all the
physical
quantities. The case of a magnetic impurity coupled to a Fermi system
with a linear density of states seems to be analogous to the behavior of
critical systems at their upper critical dimension.
In contrast, the case of the constant density of states is the analog of a
critical
system at the lower critical dimension (marginal instability).
The closest analog to this problem is the critical behavior of the Gross-Neveu
Model in
$3+1$ (space-time) dimensions~\cite{kocic}.

The results that we find here hold for the simple regime of large $N_c$ which
does not
describe correctly the dynamics of the system with more than one fermion
flavor.
At this leading order in $1/N_c$, the existence of several channels is
 only reflected into a trivial degeneracy and it does not lead to
any new physics. However, as has been shown by Blandin and
Nozi{\`e}res~\cite{nozieres}, already the leading corrections in $1/N_c$ will
lead to non-trivial behavior in the ``screened" phase, $g_0 > g_c$.
In the standard Kondo problem, this behavior was confirmed by both the
Bethe-Ansatz
solution~\cite{multichannel} and Conformal Field Theory~\cite{affleck-ludwig}.
We note here that, due to the non-local character of the effective
one-dimensional
theory discussed in section \ref{sec:toy}, it is not possible to use neither
the
Bethe-Ansatz nor Conformal Field Theory to study the non-critical Kondo problem
that we discuss here. Nevertheless, we expect that the physics of the phase
with
$g_0>g_c$ should be similar to that of the conventional overscreend Kondo
problem.
We also note that, in a more realistic model, one expects {\it four} nodes
with two
nodes of each chirality and, hence with more flavors and a more pronounced
overscreened
behavior.

We close this section with a discussion of the zero-field susceptibility for
this
model. The coupling of a magnetic field only
to the impurity spin has the form of a Zeeman term
\begin{equation}
H_h  = -  \sum_{\sigma=1}^{N_c} f^{\dagger}_\sigma {\hat h}_\sigma f_\sigma
\label{eq:magsource}
\end{equation}
Here, ${\hat h}_\sigma$ are the elements of a suitably chosen $SU(N_c)$
diagonal
generator, which also includes the appropriate normalization factors so as to
set
the Bohr magneton to one. We will need to ensure the tracelessness of the
generator,
and we can do so by keeping all the elements of order one with alternating
signs.
In this way we prevent the case of overpopulating one of the states that the
Zeeman term splits with respect to the others, {\it i.e.,\/} we explicitly
avoid cases like the one with $N_c-1$ elements being $-1$ and the last element
being $N_c-1$. This is an important consideration to care about if we want to
study the magnetic field crossover, {\it i.e.,\/} if we were interested in
obtaining the suceptibility at finite field. This was discussed in
reference~\cite{withoff2}. However, we will be interested
only in the zero-field suceptibility and it will be sufficient to assume all
the ${\hat h}_\sigma$ to be $\pm h$, so that the sum of their squares gives a
factor of $N_c$.
The magnetization $M$ is given by
\begin{equation}
 M  =  \sum_{\sigma=1}^{N_c} \langle f^{\dagger}_\sigma {\hat h}_\sigma
f_\sigma
\rangle =
 -  \sum_{\sigma=1}^{N_c} \int_{-\infty}^{\infty} \frac{d\omega}{2\pi}
\frac{{\hat h}_\sigma}{\epsilon_f - i\omega\left(1 - N_f|\varphi_0|^2
f(\omega^2,
\Lambda^2)
\right) - {\hat h}_\sigma}
\label{eq:magnet}
\end{equation}
The susceptibility at zero field  is
\begin{equation}
\left. \chi \right|_{h=0}  = \left. {\frac{\partial M}{\partial
h}}\right|_{h=0}
=
\frac{N_c}{\Lambda} \left\{ \frac{2}{\nu} \ - \
\frac{\partial}{\partial\nu}\right\}
\int_{0}^{\infty} \frac{dx}{\pi}
\ \frac{\nu}
{ \nu^2 + x^2 \left( 1 - \Delta \ \frac{\log x}{1-x^2} \right)^2}
\label{eq:suscept}
\end{equation}
Once again, we consider two regimes:~$Q_f/N_c \to 0$ and
$Q_f/N_c \to 1/2$. We discuss
only the regime $g_0>g_c$ since the suceptibility is infinite
below $g_c$ , where the impurity is unscreened.

\newcounter{letras2}
\begin{list}
{\alph{letras2}~)}{\usecounter{letras2}}
\item
$\frac{Q_f}{N_c}\ll 1$:\\
For $\frac{Q_f}{N_c} \ \approx \ \frac{\nu}{\pi} \ e^{1/\Delta}$,
we get
\begin{equation}
\chi(0)  \approx \frac{Q_f}{|\epsilon_f|}=\frac{Q_f}{T_K}
 \approx
\frac{N_c}{\pi v_F^2} (\frac{Q_f}{N_c})^2 \left( \frac{J_c J_0}{J_0-J_c}
\right)
\log\left[
\left(\frac{2 Q_f}{N_c}\right)^2 \left(
\frac{J_0}{J_0-J_c}
\right) \right]
\label{eq:suscb}
\end{equation}
\item
$\frac{Q_f}{N_c} \ \approx \ \frac{1}{2}$:\\
For $\frac{Q_f}{N_c} \ \approx \ \frac{1}{2} - \frac{3}{\pi \nu^3} \
e^{-3/\Delta}$, we have
\begin{equation}
\chi(0)  \approx
\frac{N_c}{2\pi v_F^2} \
\left({\frac{J_c J_0}{J_0-J_c}}\right)
  \log \left[{\frac{2}{ \pi}}\left({\frac{J_0}{J_0-J_c}}\right)\right]
\left\{ 1 - 5 \left( \frac{1}{2} - \frac{Q_f}{N_c} \right) + \ldots\right\}
\label{eq:susca}
\end{equation}
\end{list}
Once again, we find logarithmic corrections to scaling.

We see from Eq.(\ref{eq:susca}) and Eq.(\ref{eq:suscb})
that the zero-field susceptibility $\chi(0)$, at zero temperature,
has a behavior consistent with the general picture
described above. It is finite for $g_0 \ > \ g_c$, but diverges as
$g_0$ approaches $g_c$. This is consistent with the impurity being
screeened (Kondo effect) for a value of the coupling constant bigger
than the critical value. However, as we approach the critical coupling
constant from the screening (Kondo) phase, the susceptibility must diverge,
since for values
of the coupling lower than the critical value, $\Delta$ vanishes and the
impurity effectively decouples from the band. In this regime one expects the
impurity will behave like a free spin in a magnetic field and, since we are
at zero temperature, the susceptibility should diverge.

\section{T-matrix and one particle Green function}
\label{sec:prop}

In this section we derive the propagator for the band electrons to order
$N~\to~\infty$.
We will use this expression to calculate the $T$-matrix in $N \to \infty$
limit. We will
consider processes in which an electron with a given energy is initially in a
 state with
 well defined {\it angular momentum}, and it is scattered by the impurity into
 a state
with possibly a different angular momentum. Thus, we will
parametrize the $T$-matrix by the magnitude of the incoming and the outgoing
momenta
(namely, the energy of the state) and by
the angular momenta of channels involved in the scattering process. This
description is
natural since at most only a pair of angular moementum states ($\ell=0,-1$ with
our choice of
node) are actually mixed by the
impurity.

In contrast, the computation of the correlation function
$\left< \Psi^{\dagger}_{j}(\vec r)~\Psi_{k}(\vec r \ ')\right>$, for
$\vec r$ and $\vec r \ '$ away from the impurity
will involve a mixing of all the angular momentum channels.
However, in principle, we could
add a source term and also decompose the source field in angular momentum
modes,
and keep only the modes that get mixed by the interaction with the impurity.
In other words, we are considering the scattering of band electrons
among angular momentum channels produced by the interaction with the impurity.
In the case of a coupling to a flat band  (usual Kondo problem), the only
angular momentum channel that gets involved is the $s-wave$. Thus, in practice,
we only need
to consider the partial waves that actually get mixed by scattereing from the
impurity.

Therefore, it will be sufficient to work with the equivalent 1-dimensional
fermions
$c_1$ and $c_2$ while keeping in mind the correspondences defined in section
\ref{sec:toy}.
The relevant part of the action is
\begin{eqnarray}
S_F  & = &
\int_{-\infty}^{\infty} \frac{d\omega}{2\pi}
\int_{-\infty}^{\infty}
\frac{dp}{2\pi}
 c^{\dagger}_{l \sigma}(p,\omega) \ \left( -i\omega +  E(p)\right)
 c_{l \sigma}(p,\omega)
\nonumber \\
& + &
\int_{-\infty}^{\infty} \frac{d\omega}{2\pi}
\int_{-\infty}^{\infty} \frac{d\omega  '}{2\pi}
\int_{-\infty}^{\infty}
\frac{dp}{2\pi} \ \sqrt{|p|}
\Big\{ \varphi^*_l(\omega - \omega  ') f_\sigma^{\dagger}(\omega  ') c_{l
\sigma}(p,\omega)
\nonumber \\
& & \qquad \qquad + \
\varphi_l(\omega - \omega  ')
 c^{\dagger}_{l \sigma}(p,\omega)  f_\sigma(\omega  ')
\Big\}
\end{eqnarray}
We add sources $\eta^{\dagger}_{l \sigma} (p,\omega)$ and $\eta_{l
\sigma}(p,\omega)$
for the fields $c_{l \sigma}(p,\omega)$ and $c^{\dagger}_{l \sigma}$ by adding
the following source term to the action
\begin{equation}
\int_{-\infty}^{\infty} \frac{d\omega}{2\pi}
\int_{-\infty}^{\infty}
\frac{dp}{2\pi}
\left( \eta^{\dagger}_{l \sigma}(p,\omega) c_{l \sigma}(p, \omega) \ + \
c^{\dagger}_{l \sigma} (p, \omega) \eta_{l \sigma} (p,\omega)
\right)
\end{equation}
After completing squares and integrating over
the band fermion fields, we are left with (appart from a normalization factor)
an action which depends only on the impurity fermion fields and the sources,
\begin{eqnarray}
S_f & = &
\sum_{\sigma=1}^{N_c} \int \frac{d\omega}{2\pi}
f^{\dagger}_\sigma (\omega) \left( -i\omega + \epsilon_f\right) f_\sigma
(\omega)
\nonumber \\
& - &
\sum_{a, w}^{} \int_{-\infty}^{\infty} \frac{d\omega}{2\pi}
\int_{-\infty}^{\infty}
\frac{dp}{2\pi}
\left[ \int \frac{d\omega '}{2\pi} \varphi_l^* (\omega - \omega ')
\sqrt{|p|} f^{\dagger}_\sigma (\omega ') + \eta^{\dagger}_{l \sigma} (p,\omega)
\right]
\left( \frac{1}{-i\omega + E(p)}\right)
\nonumber \\
& & \kern2truein
\left[ \int \frac{d\omega ''}{2\pi} \varphi_l^* (\omega - \omega '')
\sqrt{|p|} f_\sigma (\omega '') + \eta_{l \sigma} (p,\omega) \right]
\nonumber \\
& + & \int_{-\infty}^{\infty} \frac{d\omega}{2\pi}
\int_{-\infty}^{\infty}
\frac{dp}{2\pi}
\eta^{\dagger}_{l \sigma} (p,\omega)
\left( \frac{1}{-i\omega + E(p)}\right) \eta_{l \sigma} (p,\omega)
\end{eqnarray}
After a few fairly straightforward manipulations and another integration
now over the impurity fermion fields, we can re-write the action as
\begin{equation}
S_f \left[ J, J^{\dagger} \right] \ = \
- \ \sum_{\sigma=1}^{N_c}
\int \frac{d\omega}{2\pi} \ \int \frac{d\omega '}{2\pi}
J^{\dagger}_\sigma (\omega) \ K^{-1} (\omega, \omega ') \ J_\sigma(\omega ')
\end{equation}
where we have introduced the operators $J_\sigma$'s and the kernel $K$ which
are given by
\begin{equation}
J_\sigma (\omega) \ = \ \sum_{l=1}^{N_f}
 \int \frac{d\omega '}{2\pi} \int \frac{dp}{2\pi}
\varphi^*_l(\omega ' - \omega)
\ \frac{\sqrt{|p|}}{-i \omega ' + E(p)} \
\eta_{l \sigma} (p,\omega ')
\end{equation}
and
\begin{equation}
K (\omega, \omega ') \ = \
\left( -i\omega + \epsilon_f\right) \ \delta(\omega - \omega ')
 -
\int \frac{dp}{2\pi}
\int \frac{d\omega ''}{2\pi}
\ \frac{|p|}{-i \omega ' + E(p)} \
\sum_{l=1}^{N_f}
\varphi^*_l(\omega '' - \omega) \ \varphi_l(\omega '' - \omega ')
\end{equation}
The generating functional for the Green's functions (at zero temperature) can
be written as
\begin{equation}
Z_F \left[ \eta, \eta^{\dagger}\right]
\ = \
\exp \left( - \ S_f \left( \eta, \eta^{\dagger} \right) \right)
\end{equation}
so that correlation functions of the band fermion fields can be computed by
differentiation over $Z_F$
\begin{eqnarray}
\left< c^{\dagger}_{l \sigma} (q,\Omega) \ c_{l' \sigma'} (q', \Omega ')
\right>
\ = \
\left< \ \frac{\delta}{\delta \eta_{l \sigma}(q,\Omega)}
\ \frac{\delta}{\delta \eta_{l' \sigma'} (q', \Omega')}
\ Z_F \left[\eta, \eta^{\dagger}\right] \right>_{\eta =\eta^{\dagger} = 0}
\end{eqnarray}
We restrict the calcultation to the static solution case
where $\varphi_{l} (\Omega  - \omega) \equiv {\bar \varphi}_{l} \ \delta(
\Omega  - \omega)$.
Then
\begin{eqnarray}
\left<
 c^{\dagger}_{l \sigma} (q,\Omega) \ c_{l' \sigma'} (q', \Omega ') \right> \
 = & & \
\delta_{l l'} \ \delta_{\sigma \sigma'} \ \delta(q - q') \ \delta(\Omega -
\Omega') \
\frac{1}{-i\Omega + E(q)}
\nonumber \\
+ \
\frac{\sqrt{|q|}}{-i\Omega + E(q)} & & \
{\cal T}_{l l'}(\Omega)
\frac{ \sqrt{|q'|}}{-i\Omega + E(q')} \ \delta(\Omega - \Omega ')
\end{eqnarray}
where the $T$-matrix ${\cal T}_{l l'}(\Omega)$ is given by
\begin{equation}
{\cal T}_{l l'} (\Omega)
\equiv -
\frac{{\bar\varphi}_{l} \ {\bar \varphi}^*_{l'}}{-i\Omega + \epsilon_f +g_0(i
\Omega)
 \sum_{l=1}^{N_f} |{\bar \varphi}_l|^2}
\label{eq:tmatrix}
\end{equation}
where $g_0(i \Omega)$ was given in Eq.~(\ref{eq:g0}) and Eq.~(\ref{eq:g00}).

The $T$-matrix of eq.~(\ref{eq:tmatrix}) exhibits several important features:
\newcounter{letras3}
\begin{list}
{\alph{letras3}~)}{\usecounter{letras3}}
\item
Eq.~(\ref{eq:tmatrix}) gives the $T$-matrix at {\it imaginary} frequency.
The $T$-matrix for {\it real}  frequency (at zero temperature)
is found by an analytic continuation
to the real axis, $i \Omega \to \Omega$.
\item
If  ${\bar\varphi}_{l}$ and $ {\bar \varphi}^*_{l'}$ are non-zero,
there exists a finite matrix element for the scattering of a band
electron from a state with angular momentum $l$ into a state with angular
momentum  $l'$ (and viceversa). In contrast, for a flat band, the $T$-matrix is
diagonal in
angular momentum states. This matrix element is not invariant under $SU(N_f)$
since
it mixes the angular momentum channels. As we pointed out above, this is a
consequance of the
unphysical spontaneous breaking of the flavor (channel) symmetry. Thus only the
{\it trace} of
this $T$-matrix is physical. All other matrix elements will be suppressed by
infrared
divergent corrections to next order in the $1/N_c$ expansion.
\item
{}From the analytic continuation of the trace of the $T$-matrix one can find
the position of the Kondo resonance $\Omega_K$,
the width $W_K$ and the phase shift $\delta(\Omega)$.
Using the fact that $\sum_{l=1}^{N_f}|{\bar\varphi}_l|^2=N_f |\varphi_0|^2$, we
can write the trace in the form
\begin{equation}
{\rm tr} {\hat {\cal T}}(\Omega)\equiv {\cal T}_{l l} (\Omega)
=
-{\frac{\pi v_F^2 \Delta}
{\Omega +| \epsilon_f| +\Omega \Delta \log {\frac{\Lambda}{|\Omega|}}-
i {\frac{\pi }{2}}\Omega \Delta}}
\label{eq:trace}
\end{equation}
which is valid for frequencies low compared to the bandwidth $\Omega \ll
\Lambda$ but
comparable to the dynamically generated scales.
The phase shift $\delta(\Omega)$ is
\begin{equation}
\delta(\Omega)= \tan^{-1}\left(
{\frac{{\frac{\pi }{2}}\Omega \Delta}
{\Omega + |\epsilon_f| + \Omega \Delta \log {\frac{\Lambda}{|\Omega|}}
}}\right)
\label{eq:phaseshift}
\end{equation}
\item
The imaginary part of the $T$-matrix is
\begin{equation}
{\rm Im}\; {\rm tr} {\hat {\cal T}}(\Omega)
=
\frac{\frac{1}{2} (\pi\Delta v_F)^2 \Omega}
{
\left(
|\epsilon_f| + \Omega \left(1 - \Delta \log {\frac{|\Omega| }{ \Lambda}}
\right)
\right)^2
+
\left( \frac{\pi}{2}\Delta\Omega\right)^2
}
\label{eq:imT}
\end{equation}
The function given by Eq.(\ref{eq:imT}) has a peak at the Kondo resonance
$\Omega_K$,
which is the solution of the equation
\begin{equation}
\epsilon_f = \Omega \left(1 - \Delta \log {\frac{|\Omega| }{ \Lambda}} \right)
\end{equation}
One easily obtains the asymptotic solution
\begin{equation}
\Omega_K \approx -
\pi v_F^2 {\frac{ N_c}{Q_f}} \left({\frac{J_0-J_c}{J_0 J_c}}\right)
{\frac{1}{\log\left[ \frac{2}{\pi} \left({\frac{2 Q_f}{N_c}}\right)^3
\left({\frac{J_0}{J_0-J_c}}\right)^2 \right]}}+\ldots
\label{eq:omegaK}
\end{equation}
where we have kept only the leading logarithmic corrections.
We see immediately that we have a resonance centered at $\Omega_K$, which
is basically the Kondo scale $\epsilon_f$ that we found in section
\ref{sec:scaling}
but modified by the ubiquitous logarithmic corrections.
In particular, the position of the Kondo resonance scales with the distance to
the
critical coupling constant and it goes to zero (the Fermi energy) at $J_c$.
\item
The resonance width $W_K$ can be read off from Eq.(\ref{eq:imT})
\begin{equation}
W_K  \approx |\Omega_K|~\frac{\pi \Delta}{2}
\label{eq:WK}
\end{equation}
{}From Eq.~(\ref{eq:phaseshift}) it is clear that the dimensionless parameter
$\Delta = N_f |{\bar\varphi}_0|^2/\pi v_F^2$
sets the scale for the range of the scattering amplitude.
Thus, as far as transport properties such as the resistivity are concerned,
$\Delta$ is the important parameter,
whereas thermodynamic properties such as the suceptibility,
depend entirely on the impurity energy scale $\epsilon_f$, {\it i.e.,\/}
the Kondo scale $T_K$. The analysis done in the last section shows that,
due to the strong logarithmic corrections, these two  scales are  completely
different.
Once again, in the conventional Kondo effect, $T_K$ controls both effects.
\end{list}
Hence, in contrast with conventional Kondo behavior, the interaction with the
magnetic
impurity induces a non-trivial renormalization of the
propagator of the band fermions. This will happen even for $J_0 \leq J_c$,
although
this effect only occurs in higher order in $1/N_c$. Finally
we stress that, due to the $SU(N_f)$ flavor symmetry,
there cannot be any channel
mixing scattering processes in this model.
However, in the following
section, we will present a model which describes the most general impurity
scattering processes for
systems with only one node. In that model the channel (or flavor) symmetry
$SU(N_f)$ is
broken explicitly already at the level of the impurity Hamiltonian. Thus we
expect
to find processes which will mix the various channels.

\section{Generalized impurity}
\label{sec:genimp}

In the past sections we considered a situation in which the fermions coupled to
the
impurity spin only through  their own  spin density $\Psi^\dagger {\vec \tau}
\Psi$.
Scattering processes of this type have the simplifying feature that particles
and holes
interact with the impurity independently from each other and exactly in the
same way.
However, in practice, more general scattering processes will be present.
We also made the assumption that only one node of the
two-dimensional fermions is present.

Here, we will consider all possible scattering processes involving {\it only
one node}.
The coupling through the spin density has the very special feature
that it is {\it diagonal} in the components of the spinors, {\it i.e.,\/} the
impurity
 does not mix particles with holes. In terms of the effective Hamiltonian of
Eq.~(\ref{eq:halt}), this diagonal coupling implies that
there is no {\it explicit} mixing of angular momentum channels.
Hence, the model has an $SU(N_f)$ symmetry (with $N_f=2$). Any process
which mixes particles and holes will break the flavor symmetry explicitly.
We will include now these processes.

We will also incorporate the correct node structure. For example,
in the case of a flux phase say,
on a square lattice, there will be four different {\it nodes} corresponding
to the symmetry points $(\pm \pi/2a_0, \pm \pi/2a_0)$ of the first Brillouin
zone.
As we discussed in section \ref{sec:intro},
a $d$-wave superconductor has an analogous node structure, at the points where
the gap has nodes. We will discuss of this very interesting
case in a separate publication \cite{next}.
It is straightforward to see that the main effect of a more general node
structure is
to {\it increase} the number of {\it flavors} in the effective model. In the
one-node
model of the previous section we found that there were two flavors and an
$SU(2)$
flavor symmetry. The existence of two flavors can be traced back to the spinor
structure
of the original problem. When the two-dimensional fermions have more than one
node, the
number of flavors becomes {\it twice the number of nodes}, $N_f=2 N_{nodes}$.
Thus,
for a flux phase we will have $N_f=8$ flavors.
However, flux phases are {\it even} under parity. Hence, there are two pairs of
nodes with opposite parity. The parity of the node is given by the {\it
relative sign}
of the two terms in the free fermion Hamiltonian of Eq.~(\ref{eq:h0}). Thus, a
change
of parity is equivalent to the mapping $\theta \to -\theta$, in other words to
a
reflection across the $x_1$ axis. In Section \ref{sec:toy} we showed that,
with the choice of parity we made there, the channels with angular momenta
$\ell=0$ and $\ell=-1$ get mixed by the impurity.
Hence, if nodes with the other parity are also included,
we will also get mixing between $\ell=0$ and  $\ell=+1$,
but no direct mixing between $\ell=+1$ and $\ell= -1$.
In Section \ref{sec:toy} we also found that only the mapping between
the components of the original fermion and the fermions of the effective
one-dimensional theory carried information about which angular momentum
components are mixed. This feature is always present.
Thus, the net effect of having a multinode structure is to increase the
number of flavors. Consequently, although the thermodynamic properties will be
insensitive to parity assignments, the scattering amplitudes of the original
fermions with the impurity will carry the information of the parity of the
nodes.
This feature can be used to determine the nature of a superconducting state.

In what follows, we will consider  a system with four nodes, two with positive
parity
and two with negative parity, interacting with a single magnetic impurity.
For simplicity we will assume that the impurity form factor is strongly peaked
at small momentum, so that inter-node scattering processes can be ignored.
We will keep, however, processes in which particles and holes may interact
differently
with the impurity and/or mix with each other.
We parametrize these processes with four coupling
constants $J_i$, with $i=0,1,2,3$.
The generalized impurity Hamiltonian is given by
\begin{equation}
H_{imp} = \sum_{l,i} J_i \; \Psi^{\dagger}_{\alpha l}(0)\
 T_i \ {\vec \tau} \ \Psi_{\beta l}(0) \ {\vec S}
\label{eq:gen}
\end{equation}
where the indices $l$ labels the nodes and $\alpha$ and $\beta$ the spin
components.
$T_0=I$ is the $2 \times 2$ identity matrix and $T_i=\sigma_i$ ($i=1,2,3$) are
the
three Pauli matrices. Here $J_0$ is the coupling constant that was
used in the model of Section~\ref{sec:toy}.

The free fermion Hamiltonian for each node
\begin{equation}
H_0 = \int \frac{d^2p}{(2\pi)^2} \Psi^{\dagger}(p) v_F {\vec \sigma}
\cdot {\vec p} \Psi(p)
\nonumber
\end{equation}
is invariant under the symmetry transformation
${\rm C} \Psi(p) \to  \sigma_3 \Psi (-p)$.
The effect of this transformation on on $H_{\rm imp}$ is to reverse the sign
of both $J_1$ and $J_2$. This implies that the sign of $J_1$ and $J_2$
is irrelevant to the properties of the theory, which should depend
only on the absolute values of these two coupling constants.
In the expressions for the Fermi field
\begin{equation}
\Psi_{i l}(0) \ = \ \frac{1}{\sqrt{2}}
\int_{-\infty}^{\infty} \frac{dp}{2\pi} \ \sqrt{|p|} \ c_{i l}(p)
\qquad \qquad {\rm and} \qquad \qquad
S^a \ \equiv \ f_{\gamma}^{\dagger} \tau^a_{\gamma\delta} f_{\delta}
\end{equation}
we have $l=1, \ldots, N_{nodes}$ and $i=1,2$. In what follows we use a single
flavor
index $i=1, \ldots, N_f$ with $N_f=2 N_{nodes}$.
However, we will keep in mind that the impurity does not mix
different nodes but it does mix the flavor components associated with the same
node.

The model of equivalent one-dimensional right-moving excitations has an $SU(2)$
spin (color) symmetry and an $SU(2 N_{nodes})$ flavor symmetry
(associated with the channels) broken down to $SU(N_{nodes})$.
The free fermion kinetic term has the form
\begin{equation}
H_0 \ = \ \int \frac{dp}{2\pi} \ pv_F \sum_{i \alpha} c_{i
 \alpha}^{\dagger}(p)
c_{i  \alpha}(p)
\end{equation}
and an impurity interaction term with the following structure
\begin{equation}
H_{imp} = \  -  \int \frac{dp}{2\pi} \int \frac{dq}{2\pi} \sqrt{|p||q|}
f_\alpha^{\dagger} c_{\alpha j }(q) c^{\dagger}_{ \beta i }(p) f_\beta    T_{i
j}
\label{eq:himp}
\end{equation}
where $T_{i j}$ is the coupling matrix which has a block diagonal form.
For $i,j$ associated with the same node $T_{ij}$ has the form
\begin{equation}
T_{i j} \equiv N_c \left( J_0 \delta_{i j} + J_1 \sigma_1^{i j}
+ J_2 \sigma_2^{i j} + J_3 \sigma_3^{i j} \right)
\label{eq:Tij}
\end{equation}
In other terms the matrix has the form $T \otimes I$ where $T$ is the $2 \times
2$
matrix of Eq.~(\ref{eq:Tij}) and $I$ is the $N_{nodes} \times N_{nodes}$
identity matrix. We have also let the spin indices $\alpha,\beta$ to run from
$1, \ldots, N_c$ and used the identity of Eq.~(\ref{eq:suntrace}).
The form of the impurity Hamiltonian Eq.~(\ref{eq:himp}) shows that,
if the coupling constants $J_i$ are all different,
the $SU(2)$ flavor symmetry of each node is broken by the interactions
but the symmetry involving different nodes remains intact.

We will now proceed as in the previous sections and solve this model in the
limit of $N_c \to \infty$. After a Hubbard-Stratonovich transformation,
the quartic term of the euclidean action becomes (repeated indices are summed)
\begin{equation}
\int d\tau
\sigma_{i}^{*}(\tau)  M_{ij} \sigma_{j}(\tau)
 +  \int d\tau  \int_{-\infty}^{\infty} \frac{dp}{2\pi}
\sqrt{|p|} \left(
\sigma_{i} (\tau) \ f^{\dagger}_{\alpha}(\tau) c_{\alpha i}(p,\tau)
+  \sigma^*_{i}(\tau)  \ c^{\dagger}_{\alpha i}(p, \tau) f_{\alpha}(\tau)
\right)
\label{eq:HSgen}
\end{equation}
where $ M_{\alpha\beta}  =  (T_{\alpha\beta})^{-1}$.
Since $T$ is of the form $T \otimes I$, then $M$ has the same form,
{\it i.e.,\/} $M \otimes I$.
In the following, $M$ stands for the $2\times 2$ matrix of
Eq.~(\ref{eq:HSgen}).
It can be easily shown that
\begin{equation}
M \ = \ \frac{N_c}{g_0^2 - {\vec g}\cdot{\vec g}} \left(
\begin{array}{lr}
 g_0 \ - \ g_3 & - \ (g_1 \ - \ ig_2) \\
            - \ (g_1 \ + \ ig_2) & g_0 \ + \ g_3
\end{array} \right)
 \equiv N_c \ {\tilde M}
\end{equation}
where $J_i \ \equiv \ g_i/N_c^2$ and ${\vec g} \equiv (g_1, g_2, g_3)$.
In the case considered in section \ref{sec:toy}, {\it i.e.,\/}
when $J_i = 0$, $J_0 \neq 0$, we have a full $U(2)$ flavor
symmetry between the $\sigma$-fields. In other words, the two channels play
exactly the same role. When the coupling constants are all different,
this symmetry is broken down to a $U(1)\times U(1)$, where one of the $U(1)$
symmetries is generated by the identity and the other by $\tilde M$.

\begin{figure}[htb]
     \begin{center}
      \setlength{\unitlength}{1cm}
       \begin{picture}(14,18.5)           % figure dimensions
        \put(-3.5,-7){\includegraphics{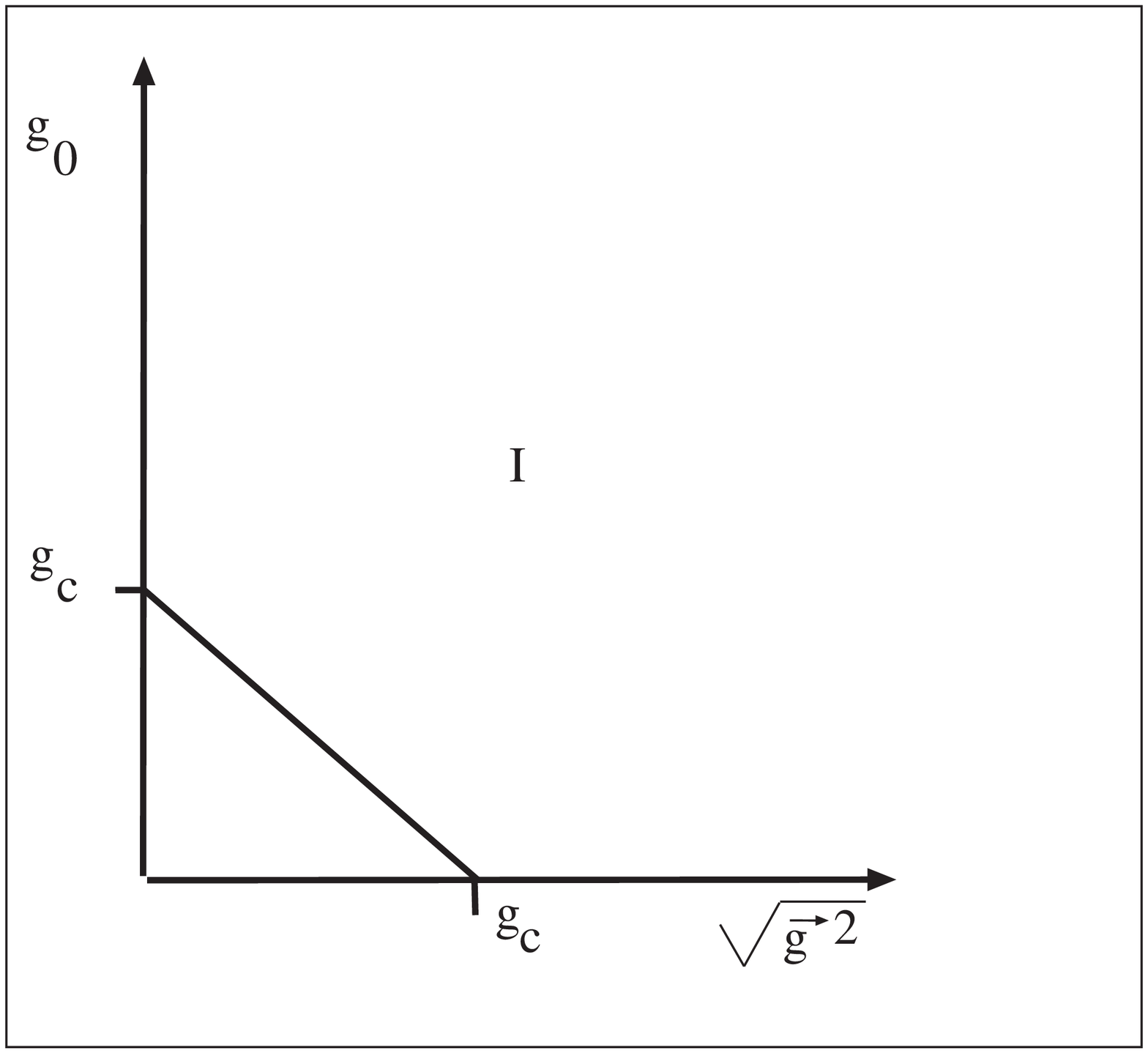}}
       \end{picture}
      \end{center}
\caption{Phase diagram of the generalized impurity model.}
\label{figure}
\end{figure}

The $N_c \to \infty$ limit is taken in the standard fashion. The
Hubbard-Stratonovich
fields $\sigma_i$ will be chosen to be an arbitrary vector for each node
and the same vector for all nodes. Let $V$ be a unitary transformation which
diagonalizes the $2 \times 2$ matrix $\tilde M$, and
let $\sigma \equiv V{\tilde \sigma}$, so that
$|\sigma_1|^2 + |\sigma_2|^2  = |{\tilde \sigma}_1|^2 + |{\tilde\sigma}_2|^2$.
The eigenvalues of $\tilde M$ can be rewritten as $m\pm\delta m$, where
\begin{equation}
m \ = \ \frac{g_0}{g_0^2 - {\vec g}^2}
\qquad \qquad {\rm and}
\qquad \qquad \delta m \ =  \ \frac{\sqrt{{\vec g}^2}}{g_0^2 - {\vec g}^2}
\label{eq:evs}
\end{equation}
The modified effective action can be written as
\begin{eqnarray}
\frac{1}{\beta}~S_{eff} & = & -  N_c
\int_{-\infty}^{\infty} \frac{d\omega}{2\pi}
\ \log
\left(-i\omega + \epsilon_f + {\frac{1}{2}} N_{nodes}
\left( |{\tilde \sigma}_1|^2 + |{\tilde \sigma}_2|^2
\right) G_0 (i\omega)\right)
\nonumber \\
& &  \qquad \qquad + \ {\frac{1}{2}} N_{nodes}
N_c \left[\left( |{\tilde \sigma}_1|^2 + |{\tilde \sigma}_2|^2\right) \ m
 +
\left( |{\tilde \sigma}_1|^2 - |{\tilde \sigma}_2|^2 \right) \ \delta m \right]
 -  Q_f  \epsilon_f
\nonumber \\
& = & -  N_c
\int_{-\infty}^{\infty} \frac{d\omega}{2\pi}
\ \log\left(-i\omega + \epsilon_f + {\frac{1}{2}} N_{nodes}
\left( |{\tilde \sigma}_1|^2 + |{\tilde \sigma}_2|^2
\right) G_0 (i\omega)\right)
\nonumber \\
& & \qquad \qquad  + \ {\frac{1}{2}} N_{nodes}
N_c \left( \frac{|{\tilde \sigma}_1|^2}{g_0 - \sqrt{{\vec g}^2}}
 +  \frac{|{\tilde \sigma}_2|^2}{ g_0 + \sqrt{{\vec g}^2}} \right)
 - Q_f \epsilon_f
\label{eq:modeff}
\end{eqnarray}
The new Saddle Point Equations are
\begin{equation}
\frac{Q_f}{N_c} \ = \ - \ \int_{-\infty}^{\infty} \frac{d\omega}{2\pi}
\ \frac{1}{-i\omega + \epsilon_f + {\frac{1}{2}} N_{nodes}
\left( |{\tilde \sigma}_1|^2 + |{\tilde \sigma}_2|^2 \right) G_0(i\omega)}
\label{eq:Qfmod}
\end{equation}
and
\begin{equation}
\frac{{\tilde \sigma}_1}{g_0 - \sqrt{{\vec g}^2}}
\ = \
{\tilde \sigma}_1 \
\int_{-\infty}^{\infty} \frac{d\omega}{2\pi}
\ \frac{G_0(i\omega)}{-i\omega + \epsilon_f + {\frac{1}{2}} N_{nodes}
\left( |{\tilde \sigma}_1|^2 + |{\tilde \sigma}_2|^2 \right)G_0(i\omega) }
\label{eq:sigma1}
\end{equation}
and
\begin{equation}
\frac{{\tilde \sigma}_2}{g_0 + \sqrt{{\vec g}^2}}
\ = \
{\tilde \sigma}_2 \
\int_{-\infty}^{\infty} \frac{d\omega}{2\pi}
\ \frac{G_0(i\omega)}{-i\omega + \epsilon_f + {\frac{1}{2}} N_{nodes}
\left( |{\tilde \sigma}_1|^2 + |{\tilde \sigma}_2|^2 \right)G_0(i\omega)  }
\label{eq:sigma2}
\end{equation}
We now solve these new SPE's. The solution will be cast in the form of a phase
diagram which
can be plotted in a $\sqrt{{\vec g}^2}~-~g_0$ plane (see figure~\ref{figure}).
In principle, we find three different solutions of the SPE's:
\newcounter{letras4}
\begin{list}
{\alph{letras}~)}{\usecounter{letras}}
\item
${\tilde \sigma}_1  =  {\tilde\sigma}_2  =  0$.
This is the region of the phase diagram below the line $g_0 + \sqrt{{\vec g}^2}
= g_c$.
This is the weak-coupling phase. The magnetic impurity is effectively
decoupled from the band electrons. To leading order in $1/N_c$ the impurity
does not interact with the band fermions and behaves like a free magnetic
moment. Consequently the impurity spin suceptibility is infinite, there is no
resonance
and there is no Kondo screening.
\item
Both ${\tilde\sigma}_1 \neq 0$ and ${\tilde\sigma}_2 \neq 0$.  There is no
consistent solution of this form unless
${\vec g}^2 \ = \ 0$. In other words, this case is possible only if the only
non-vanishing coupling constant is $J_0$, with $J_0 \ > \ J_c$,
in which case we do have the $U(2)$ symmetry. This is the line on the axis
$g_0$ of the figure~\ref{figure} for $g_0 > g_c$;
\item
${\tilde \sigma}_2  \neq  0$, $  {\tilde\sigma}_1  =  0$. A solution of this
form satisfies the
equations
\begin{eqnarray}
\frac{1}{g_0+\sqrt{{\vec g}^2}} & =&  \int \frac{d\omega}{2\pi}
\frac{G_0(i\omega)}{ -i\omega + \epsilon_f + {\frac{1}{2}} N_{nodes}  |{\tilde
\sigma}_2|^2
G_0(i\omega)}
\nonumber\\
\frac{Q_f}{N_c} & = & - \int \frac{d\omega}{2\pi}
\frac{1}{ -i\omega + \epsilon_f +  {\frac{1}{2}} N_{nodes} |{\tilde
\sigma}_2|^2
G_0(i\omega)}
\label{eq:SPEc}
\end{eqnarray}
The action of this solution is given by
\begin{eqnarray}
\frac{1}{\beta}~S_{imp}  =
-  N_c \int \frac{d\omega}{2\pi} \ \log
\left( -i\omega + \epsilon_f + N_{nodes}  \frac{|{\tilde \sigma}_2|^2}{2}
G_0(i\omega)\right)
 +
\frac{N_c |{\tilde \sigma}_2|^2}{g_0 + \sqrt{{\vec g}^2}}
 -  Q_f~\epsilon_f
\label{eq:action-c}
\end{eqnarray}
For a solution of this type, $\sigma_1$ and $\sigma_2$ are determined by
${\tilde\sigma}_1$ through
the relations
\begin{equation}
\sigma_1 \ = \ {\tilde\sigma}_2 \ \frac{\left( g_1 - i g_2\right)}
{\sqrt{2\sqrt{{\vec g}^2}\left(\sqrt{{\vec g}^2}+g_3\right)}}
\qquad \qquad {\rm and} \qquad \qquad
\sigma_2 \ = \ - \ {\tilde\sigma}_2 \ \sqrt{\frac{\sqrt{{\vec g}^2}+g_3}
{2\sqrt{{\vec g}^2}}}
\label{eq:sigmas-c}
\end{equation}
This solution is only allowed above the line $g_0 + \sqrt{{\vec g}^2} = g_c$ in
the phase diagram. This is the Kondo or screening phase.
The results of the previous sections apply to this phase.
\item
${\tilde \sigma}_2  \neq  0$, $ {\tilde\sigma}_1  =  0$. The SPE's also allow
for
solutions of this form. In principle such solutions are allowed above the line
$g_0 - \sqrt{{\vec g}^2} = g_c$. The SPE's for this solution are exactly the
same
ones found in (c) with $g_0+\sqrt{{\vec g}^2}$ replaced by $g_0-\sqrt{{\vec
g}^2}$.
Clearly this solution competes with the solution of case (c) over a significant
region
of the phase diagram. In fact its allowed region is completely included within
the allowed region for (c).
However, we argue that this solution is always metastable and never occurs.
 In fact, for any finite value of $\sqrt{{\vec g}^2}$, the interaction
between the impurity and the band electrons will be stronger in case (c) since
$g_0+\sqrt{{\vec g}^2} > g_0 - \sqrt{{\vec g}^2}$,
favoring the first case in much the same way in which the bonding state is
preferred to
the antibonding one when the degeneracy in a two-level system is lifted by a
perturbation. It is easy to check that this is indeed correct by looking for
solutions
of both types in the vicinity of the phase transition, for $g_0$ close to $g_c$
and
$\sqrt{{\vec g}^2}$ small. In addition, since there is no actual symmetry
change
between these two ``phases", we do not expect a phase transition.
\end{list}
We conclude that the generalized impurity model has just two phases:  a phase
with an unscreened impurity and a phase with a Kondo effect. The phase boundary
is at
the line $g_0 + \sqrt{{\vec g}^2} = g_c$ . The physics of the Kondo phase is
almost
identical to what was described in the previous sections. In fact, at the level
of the $N \to \infty$ theory, the physical observables of the generalized
impurity
model can be calculated using the formulas for the single node model of the
previous
sections. The main difference is that, since the $U(2)$ flavor symmetry at each
node
is now broken by the explicit form of the impurity Hamiltonian,
operators with non-trivial matrix elements in that sector are now allowed.
In particular, there will be non-vanishing, finite,  off-diagonal matrix
elements
of the band fermion $T$-matrix. In other words, the impurity will mix band
fermion
states with different angular momentum. In contrast, processes which mix
different
nodes are still strictly forbidden. Finally we note that scattering processes
of the type described here effectively reduce the number of independent
channels.
If it were not for the existence of several nodes, we would expect to find a
Kondo
effect in this phase with a completely screened impurity. Once again, the
existence of
additional channels associated with the multinode structure turns this into a
multichannel Kondo system if inter-node scattering processes are not allowed.

\section{Discussion and Conclusions}
\label{sec:conc}

In this paper we have considered the problem of flux-phase fermions coupled
locally to a single magnetic impurity. This is formally the same as the problem
of fermions with a relativistic dispersion in two space dimensions couped to a
local spin. We derived explicitely an effective theory in one-space dimension.

The physics of the effective one-dimensional theory for this problem differs
from
the conventional radial picture of the Kondo problem in several important ways.
Firstly, it is always a multichannel system. The channels reflect the spinor
structure
of the nodes and  the multiplicity of nodes. Secondly, the relativistic
dispersion
implies a density of states which vanishes linearly at the Fermi energy.
Consequently, unlike the conventional Kondo model, the effective
one-dimensional
problem has a non-local coupling between the effective right movers and the
impurity.
This is the feature that drives the Kondo effect away from marginality
and it is responsible for the  phase transition between an unscreened
impurity phase and a phase with a Kondo effect.
This is consistent with earlier results of Withoff and Fradkin.

However, for the case of  the linearly vanishing density of states that we
discussed in this paper, the nature of the scaling in the vicinity of this
zero-temperature phase transition is drastically changed. We find that all
physical quantities exhibit very simple scaling laws modified by logarithmic
corrections. We found this behavior in the Kondo scale and in the impurity spin
suceptibility. This behavior is strongly reminscent of critical phenomena at an
upper critical dimension. One important consequence of the logarithmic
corrections
is that there are more dynamical scales and that all physical quantities are no
longer
controlled by the Kondo scale $T_K$ alone.
This is particularly clear if one compares the amplitude of the $T$-matrix and
the position and width of the Kondo resonance.

The model we studied and solved here using large-$N$ methods is an interesting
problem on its own right. In a subsequent publication we will report on a study
of
a similar model for magnetic impurities in $d$-wave superconductors where we
will
draw heavily on the ideas that we developed here.

It is interesting to compare the effective one-dimensional model that we
derived
here with the conventional one-dimensional models for the conventional Kondo
problem.
The standard Kondo problem is equivalent to a model in one space dimension
with a single right mover which interacts locally with the impurity spin
through
the fermion spin density. This coupling through a density is crucial for the
physics of the Kondo problem to be correctly described by the model of
right moving fermions. The fact that the fermions are chiral (namely,
only right movers are present) means that , up to a Fermi velocity,
the fermion density and current are the same observable.
This model can be described entirely in terms of a conserved current.
This is the starting point of the Conformal Field Theory approach of Affleck
and
Ludwig\cite{affleck-ludwig}.
However, it is also crucial for the success of the approach of Anderson, Yuval
and
Hamman\cite{ayh}. In fact, the equivalence that exists between Kondo systems
and
problems of Macroscopic Quantum Coherence\cite{leggett} rely heavily on
bosonization
of models of fermion coupled locally to impurities through a density.
The fact that this density is associated to a conserved current means that
it cannot acquire anomalous dimensions. Thus, the fixed points of the
conventional
Kondo contains only marginal operators which are made marginally relevant by
quantum fluctuations.

{}From this analysis it is clear that it is not possible to describe the phase
transition that we discuss here in terms of a Conformal Field Theory coupled to
local boundary operators representing the impurity.
The non-locality of the effective one-dimensional theory is essential.
It is because the model is non-local that the operator that couples to the
impurity
can (and does) acquire an anomalous dimension.
This is the mechanism which drives the phase transition.
These models are not equivalent to any standard Macroscopic Quantum Tunneling
(MQT)
model. It may appear that, because the density of states vansihes like a power
of
the energy, these models could be related to a subohmic MQT system
which are known not to have phase transitions.
However, subohmic MQT models of quantum impurities coupled to a macroscopic
{\it bosonic} system, with subohmic spectral density. The models that we
discussed here are fermionic and are  not equivalent to a subohmic bosonic
theory.

\section{Acknowledgements}

We are grateful to K. Ingersent for useful discussions and for
bringing to our attention the possible physical scenario discussed at the end of
Sec.~\ref{sec:intro}.
This work was supported in part by NSF grants No.~NSF DMR94-24511
at the Department of Physics of the University of Illinois at Urbana-Champaign,
and DMR89-20538/24 at the Materials Research Laboratory of the
University of Illinois.

\newpage

\appendix
\section{}
\label{sec:A}

In section \ref{sec:scaling} and from Eq.(\ref{eq:scaling9}) we have
\begin{equation}
I_1 \ = \
\frac{1}{\pi} \int_0^{e^{-1/\Delta}}
dx \ \frac{\nu}{\nu^2 + x^2\left(1-\Delta \ \frac{\log x}{1-x^2}\right)^2}
\label{eq:app1}
\end{equation}
and
\begin{equation}
I_2 \ = \
\frac{1}{\pi} \int_{e^{-1/\Delta}}^{\infty}
dx \ \frac{\nu}{\nu^2 + x^2\left(1-\Delta \ \frac{\log x}{1-x^2}\right)^2}
\label{eq:app2}
\end{equation}
In the region of integration where $x >> e^{-1/\Delta}$,
$I_2$ can be approximated by
\begin{equation}
I_2 \approx \frac{1}{2} \ - \ \frac{1}{\pi} \arctan
\left( \frac{1}{\nu} e^{-1/\Delta}\right)
\label{eq:app3}
\end{equation}
On the other hand, when we are in the regime $x << e^{-1/\Delta}$,
\begin{equation}
I_1 \approx \frac{\nu}{\pi \Delta^2}
\int_0^{e^{-1/\Delta}}
\frac{dx}{\left(\frac{\nu}{\Delta}\right)^2 + x^2 \log^2 x} \ \equiv \ R(\nu,
\Delta)
\label{eq:app4}
\end{equation}
The following inequality holds
\begin{equation}
\frac{1}{\pi} \frac{\nu e^{-1/\Delta}}{\nu^2 + e^{-2/\Delta}} \
\leq \ \ R(\nu, \Delta) \ \ \leq \ \frac{1}{\pi\nu} \ e^{-1/\Delta}
\label{eq:app5}
\end{equation}
We will be interested in the following cases.
\begin{equation}
{\rm If} \ \ \ \ \ \nu \ >> \ e^{-1/\Delta}, \ \ \ \ \ {\rm we \ \ have} \ \ \
\
\ R(\nu, \Delta) \ \to \ \frac{1}{\pi\nu} \ e^{-1/\Delta} \ << \ 1.
\label{eq:app6}
\end{equation}
\begin{equation}
{\rm If} \ \ \ \ \ \nu \ << \ e^{-1/\Delta}, \ \ \ \ \ {\rm we \ \ have} \ \ \
\
\ R(\nu, \Delta) \ \to \ \frac{\nu}{\pi} \ e^{1/\Delta} \ << \ 1.
\label{eq:app7}
\end{equation}
For \ $1 \ >> \ \nu \ >> \ e^{-1/\Delta}$, we can expand the denominator in
Eq.(\ref{eq:app4}) to get
\begin{equation}
R(\nu, \Delta) \ \approx \
\frac{1}{\pi\nu} \left\{ e^{-1/\Delta}
\ - \ \frac{3}{\Delta^2} \left(\frac{\Delta}{\nu}\right)^2 e^{-3/\Delta}
 + \dots
\right\}
\label{eq:app8}
\end{equation}
We can now get the behavior of Eq.(\ref{eq:scaling1}) in these two different
regimes:
for $ e^{-1/\Delta}  <<  \nu  <<  1 $,
\begin{equation}
\frac{ Q_f}{N_c} \approx \frac{1}{2}  -  \frac{3}{\pi \nu^3} \ e^{-3/\Delta}
\ \ {\rm and} \ \
\nu \ \approx \ \left(\frac{3}{\pi}\right)^{1/3}
\ \frac{e^{-1/\Delta}}{\left( - \frac{Q_f}{N_c} + \frac{1}{2}\right)^{1/3}}
\ >> \ e^{-1/\Delta} \ \ {\rm if} \ \ \frac{Q_f}{N_c} \ \to \ \frac{1}{2}.
\label{eq:app9}
\end{equation}
the other limit corresponds to $ \nu << e^{-1/\Delta} << 1$; consequently the
contribution from $I_2$ is neglegible and
\begin{equation}
\frac{Q_f}{N_c} \approx \frac{\nu}{\pi} \ e^{1/\Delta} \ \ {\rm or}
\ \ \nu \approx \pi \frac{Q_f}{N_c} e^{-1/\Delta} << e^{-1/\Delta} \ \ \
\ \ {\rm if} \ \ \ \ \ \frac{Q_f}{N_c} << \frac{1}{\pi}
\label{eq:app10}
\end{equation}
Now we want to work out similar approximations for the other SPE.
We have
\begin{eqnarray}
I_1' & \equiv & - \ \int_0^{e^{-1/\Delta}}
dx \ \frac{\log x}{1-x^2} \left( 1 - \Delta \ \frac{\log x}{1-x^2}
\right)
\ \frac{x^2}{\nu^2 + x^2\left( 1 - \Delta \
\frac{\log x}{1-x^2}
\right)^2}
\nonumber \\
& \approx &
\frac{1}{\Delta} \ e^{-1/\Delta} \ - \ \frac{\pi\nu}{\Delta} \ R(\nu, \Delta)
\label{eq:app11}
\end{eqnarray}
in which we made use of the fact that over the integration interval, the
variable $x$ is very small, so that the approximations $1-x^2 \approx 1$
and $-\Delta \log x >> 1$ are consistent.
For the other portion of the integral ({\it i.e.,\/} for $I_2'$) we are in
a situation where $x > e^{-1/\Delta}$ and then
$- \ \Delta \ \frac{\log x}{1-x^2} < 1$, so
we neglect the logaritmic term against 1.

We need here the following result
\begin{equation}
\int_0^{\infty} dx \ \frac{x^2 \ \log x}{1-x^2} \ \frac{1}{\nu^2+ x^2}
\ = \ \frac{1}{2} \ \frac{1}{1+\nu^2} \ \left\{ \frac{\pi^2}{2} + \pi \nu
\log\nu
\right\}
\label{eq:app12}
\end{equation}
Then
\begin{eqnarray}
I_2' & \equiv & - \ \int_{e^{-1/\Delta}}^{\infty}
dx \ \frac{\log x}{1-x^2} \left( 1 - \Delta \ \frac{\log x}{1-x^2}
\right)
\ \frac{x^2}{\nu^2 + x^2\left( 1 - \Delta \
\frac{\log x}{1-x^2}
\right)^2}
\nonumber \\
& \approx &
- \int_{e^{-1/\Delta}}^{\infty} dx \ \frac{x^2 \log x}{1-x^2} \
\frac{1}{\nu^2+x^2}
\nonumber \\
& \approx &
\frac{1}{2} \ \frac{1}{1+\nu^2} \left\{ \frac{\pi^2}{2} + \pi\nu \log \nu
\right\}
- \ e^{-1/\Delta} \left( 1 + \frac{1}{\Delta} \right)
\nonumber \\
& & \qquad \qquad \qquad \ \ \
- \ \nu \log \nu \ \arctan\left(\frac{1}{\nu} e^{-1/\Delta} \right)
- \ \nu \ {\rm F}\left(\frac{1}{\nu} e^{-1/\Delta}\right)
\label{eq:app13}
\end{eqnarray}
The function ${\rm F}(z)$ verifies that, for $|z| \ > \ 1$, ${\rm F}(z) \ = \
{\rm F}\left(\frac{1}{z}\right)$ and for $|z| \ < \ 1$ is given by
\begin{eqnarray}
{\rm F}(z) \ & = & \ \sum_{n=0}^{\infty} (-1)^n \ \frac{1}{2n+1} \ z^{2n+1}
\ \left[ \log z \ - \ \frac{1}{2n+1} \right]
\nonumber \\
& & \qquad \qquad \approx \ z \left( \log z - 1 \right)
- \frac{1}{3} \ z^3\left( \log z - \frac{1}{3}
\right) + \dots
\label{eq:app14}
\end{eqnarray}
Getting everything together,  in the limit $ \nu \gg e^{-1/\Delta}$
and $\nu, \Delta \ << 1$, where
\[
R(\nu, \Delta) \ \to \ \frac{1}{\pi\nu} \ e^{-1/\Delta}; \]
one can write the SPE in the form
\begin{equation}
\frac{1}{g_0} \ \approx \
\frac{\pi^2}{4} \ + \ \frac{\pi}{2} \ \nu \log \nu
\end{equation}
which can be solved by iteration (see Eq.(\ref{eq:nu1}) in section
\ref{sec:scaling}).
The other limiting case corresponds to $\nu \ e^{1/\Delta} << 1$, with $\nu$,
$\Delta \ << \ 1$. Here
\begin{equation}
R\left(\nu, \Delta\right) \ \to \ \frac{\nu}{\pi} e^{1/\Delta}
\end{equation}
The SPE now is
\begin{equation}
\frac{\pi^2}{4} \ - \ \frac{1}{g_0}
\approx
e^{-1/\Delta}
\left[
1 + \frac{1}{\Delta} \left(\pi \frac{Q_f}{N_c}\right)^2
+ \left( \pi \frac{Q_f}{N_c}\right)^3
+ \left( \pi \frac{Q_f}{N_c}\right)^2 \ \log\left(\pi \frac{Q_f}{N_c}\right)
\right]
\label{eq:applast}
\end{equation}
since in this case, $\nu e^{1/\Delta} \approx  \pi \frac{Q_f}{N_c}$.
The cubic term can be dropped rightaway. As for the other terms,
as we approach the transition, $\pi \frac{Q_f}{N_c}$
remains fixed while $\Delta \ \to \ 0$. Therefore, the leading term
is going to be the one with the factor $\frac{1}{\Delta}$.
This immediately gives the implicit expression for $\Delta$ in terms
of the impurity occupancy $\frac{Q_f}{N_c}$ and the distance to the
critical point $\frac{1}{g_c} - \frac{1}{g_0}$ that we
used in Eq.(\ref{eq:nu3}) in section \ref{sec:scaling}.

\begin{references}
\bibitem{foot} The usual logarithmic divergencies of the high
temperature expansion of the standard Kondo problem are absent for this
non-marginal
system. The logarithmic corrections that we discuss have a completely different
and new character.
\end{references}
\end{document}